\documentclass[fleqn,usenatbib]{mnras}

\usepackage{newtxtext,newtxmath}

\usepackage[T1]{fontenc}

\DeclareRobustCommand{\VAN}[3]{#2}
\let\VANthebibliography\thebibliography
\def\thebibliography{\DeclareRobustCommand{\VAN}[3]{##3}\VANthebibliography}

\usepackage{graphicx}	
\usepackage{amsmath}	
\usepackage{multicol}
\usepackage[dvipsnames]{xcolor}

\newcommand{\Teff}{\mbox{$T_{\mathrm{eff}}$}}
\newcommand{\Ion}[2]{#1\,\textsc{#2}}
\newcommand{\logX}[1]{\mbox{$\log(\mathrm{#1/He})$}}
\newcommand{\logXY}[2]{\mbox{$\log(\mathrm{#1/#2})$}}
\newcommand{\Msun}{\mbox{$\mathrm{M}_{\odot}$}}
\newcommand{\masy}{mas\,yr$^{-1}$}
\newcommand{\kms}{km\,s$^{-1}$}
\newcommand{\HST}{\textit{HST}}

\newcommand{\sdss}[3]{SDSS\,J#1$#2$#3}
\newcommand{\DZa}{\sdss{0956}{+}{5912}}
\newcommand{\DZb}{\sdss{1038}{-}{0036}}
\newcommand{\DZc}{\sdss{1535}{+}{1247}}
\newcommand{\DZalong}{\sdss{095645.12}{+}{591240.7}}
\newcommand{\DZblong}{\sdss{103809.09}{-}{003622.2}}
\newcommand{\DZclong}{\sdss{153505.81}{+}{124745.2}}

\title[Spectral analysis of cool DZ white dwarfs]{Spectral analysis of cool white dwarfs accreting from planetary systems: from the UV to the optical}

\author[M.A. Hollands]{
M. A. Hollands,$^{1,2}$\thanks{E-mail: M.Hollands@warwick.ac.uk}
P.-E. Tremblay,$^{2}$
B. T. G\"ansicke,$^{2}$
and D. Koester$^{3}$
\\
$^{1}$ Department of Physics and Astronomy, University of Sheffield, Sheffield, S3 7RH, UK \\
$^{2}$ Department of Physics, University of Warwick, Coventry CV4 7AL, UK \\
$^{3}$ Institut f\"ur Theoretische Physik und Astrophysik, University of Kiel,
24098 Kiel, Germany\\
}

\date{Accepted 2021 December 16. Received 2021 December 16; in original form 2021 July 22}

\pubyear{2021}

\begin{document}
\label{firstpage}
\pagerange{\pageref{firstpage}--\pageref{lastpage}}
\maketitle

\begin{abstract}
The accretion of planetary debris into the atmospheres of white dwarfs leads to
the presence of metal lines in their spectra. Cool metal-rich white dwarfs,
which left the main-sequence many Gyr ago, allow the study of the remnants of
the oldest planetary systems. Despite their low effective temperatures (\Teff),
a non-neglible amount of their flux is emitted in the near ultraviolet (NUV),
where many overlapping metal lines can potentially be detected. We have
observed three metal-rich cool white dwarfs with the Space Telescope Imaging
Spectrograph (STIS) onboard the \textit{Hubble Space Telescope} (\HST), and
compare the results determined from the NUV data with those previously derived
from the analysis of optical spectroscopy. For two of the white dwarfs, \DZb\
and \DZc, we find reasonable agreement with our previous analysis and the new
combined fit of optical and NUV data. For the third object, \DZa, including the
STIS data leads to a ten~percent lower \Teff, though we do not identify a
convincing explanation for this discrepancy. The unusual abundances found for
\DZa\ suggest that the accreted parent-body was composed largely of water ice
and magnesium silicates, and with a mass of up to $\simeq 2\times 10^{25}$\,g.
Furthermore \DZa\ shows likely traces of atomic carbon in the NUV. While
molecular carbon is not observed in the optical, we demonstrate that the large
quantity of metals accreted by \DZa\ can suppress the C$_2$ molecular bands,
indicating that planetary accretion can convert DQ stars into DZs (and not
DQZs/DZQs).
\end{abstract}

\begin{keywords}
(stars:) white dwarfs -- stars: atmospheres -- ultraviolet: stars -- planets and satellites: composition
\end{keywords}



\section{Introduction}

The majority of exoplanets orbit stars with masses $\leq 4$\,\Msun\ which are
destined to become white dwarfs, with a large fraction of their planetary
systems surviving stellar evolution
\citep{mustilletal14-1,verasetal16-1,villaver+livio07-1}. Within the Solar
System, planetary bodies beyond 1\,au are expected to survive this process
\citep{schroeder+connon08-1}, i.e. Mars, the gas giants, and the
asteroid/Kuiper belts will one day orbit a degenerate Sun
\citep{sackmannetal93-1,duncan+lissauer98-1}.

Observationally, remnant planetary systems in orbit of white dwarfs are well
established. The intense surface gravities of white dwarfs (typically $\log g
\simeq 8$, with $g$ in units of cm\,s$^{-2}$), are expected to result in
atmospheres devoid of elements heavier than hydrogen or helium, due to rapid
gravitational settling \citep{schatzman49-1,paquetteetal86-2}. The presence of
metal lines observed in the spectra of at least one quarter of white dwarfs
\citep{zuckermanetal03-1,zuckermanetal10-1,koesteretal14-1} indicates that
accretion of metal-rich material occurs frequently. Furthermore, circumstellar
dust discs are observed around some metal-rich white dwarfs due to excess infra
red flux in their spectral energy distributions
\citep{zuckerman+becklin87-1,grahametal90-1}. This led to the hypothesis that
exoplanetesimals (e.g. asteroids and minor planets) could be perturbed onto
orbits with pericentres within the white dwarf Roche radius, leading to their
tidal break up, forming a circumstellar debris disc which is then accreted onto
the white dwarf surface \citep{jura03-1}. Dynamical simulations of remnant
planetary systems indicate that planetesimals can be directed towards the
central white dwarf through gravitational encounters with other planetary
bodies \citep{debes+sigurdsson02-1,mustilletal14-1,veras+gaensicke15-1}.
Additional observational evidence comes from the presence of gas discs within
the circumstellar dust discs
\citep{gaensickeetal06-3,farihietal12-1,melisetal12-1,gentilefusilloetal21-1},
which may be related to the presence of intact solid planetesimals embedded
within some of the dusty debris discs \citep{manseretal19-1}. The fundamental
hypothesis of planetesimals reaching the tidal disruption radius
\citep{jura03-1} has unambiguously been corroborated by the detection of two
systems exhibiting deep and rapidly evolving photometric transits caused by
disintegrating planetary bodies
\citep{vanderburgetal15-1,vanderboschetal19-1,guidryetal21-1,vanderboschetal21-1}.
More recently this body of evidence has come to include bona fide planets
orbiting close to the white dwarf -- in one case detection of an evaporating
giant planet \citep{gaensickeetal19-1} and another from photometric transits of
a planet crossing the line-of-sight of the white dwarf
\citep{vanderburgetal20-1}.

Among the various observational signatures flagging the presence of evolved
planetary systems, the most common one is the presence of metal lines in white
dwarf spectra. Atmospheric modelling of these spectra, not only establishes the
stellar properties, but also the chemical composition of the accreted parent
bodies \citep{zuckermanetal07-1}. This has led to the characterisation of many
different types of planetary compositions \citep{swanetal19-1}, for example,
lithospheric material rich in Ca and Al \citep{zuckermanetal11-1}, core-like
material rich in Fe and Ni
\citep{gaensickeetal12-1,wilsonetal15-1,hollandsetal18-1}, and in one instance
\citep{xuetal17-1} cometary material rich in volatile elements such as N and S.
Some rocky material has also been established as enriched in water-ice due to
the over abundance of oxygen \citep{farihietal13-1,raddietal15-1}, and more
generally from the presence of trace hydrogen in helium-dominated atmospheres
\citep{gentileetal17-1,hoskinetal20-1,izquierdoetal21-1}. Most recently,
atmospheres enriched in the alkali metals, Li and K, have been identified in a
handful of cool white consistent with either accretion of primordial
Li-enhancement \citep{kaiseretal20-1} or continental-crust like material
\citep{hollandsetal21-1}. Additionally, Be has been detected at two systems as
a signature of spallation by high energy particles \citep{kleinetal21-1}.

In this work we focus on a specific class of metal-polluted white dwarf, the
DZs, which are distinguished by only having metal lines in their spectra. For
\Teff\ below $\approx10\,000$\,K, the strength of helium lines becomes
negligible, and so most DZs have helium dominated atmospheres, though below
$\approx5000$\,K the hydrogen Balmer series also becomes negligible in
strength, and so very cool DZs can also have hydrogen dominated atmospheres.
DZs have deep outer convection zones, which keep heavy elements suspended for
much longer than compared with the radiative atmospheres of warm DAZs. While
the corresponding diffusion timescales are on the order of Myrs, these are
still much shorter than the 1--10\,Gyr white dwarf cooling ages, thus
maintaining the requirement of accretion as the origin of the metals. On the
other hand, the Myr diffusion timescales are likely longer than the dusty
debris disc lifetime \citep{girvenetal12-1,cunninghametal21-1}, and so DZs are
rarely observed with the infrared excesses characteristic of debris discs.
Because of the low opacity of helium in the outer layers of cool DZs, their
atmospheres reach fluid-like densities at the deepest layers relevant to
atmospheric modelling \citep{blouinetal18-1}. This has the effect of making DZ
atmospheres challenging to model, as well as leading to extreme
pressure-broadening of metal lines formed at these high densities
\citep{allardetal16-1,allardetal16-2,hollandsetal17-1,blouinetal18-1,
blouinetal19-1,blouinetal19-2}.

Crucial to obtaining correct atmospheric parameters of DZs is an accurate
measurement of \Teff, as this quantity can be highly correlated with the metal
abundances. While most spectroscopic observations of DZ stars are performed at
optical wavelengths,  for some DZs (particularly the warmest), a significant
amount of flux may reside in the near-ultraviolet (NUV). Incorrect
determination of the flux level in the NUV, can therefore have a large impact
on the determined \Teff, and thus impact abundance measurements
\citep{wolffetal02-1}. To establish whether DZ atmospheric parameters can be
determined from optical data alone we have acquired NUV spectra of three
DZs\footnotemark whose optical spectra from the Sloan Digital Sky Survey (SDSS)
have previously been analysed \citep{hollandsetal17-1}: \DZalong, \DZblong, and
\DZclong\ (shortened versions used hereafter) using the Space Telescope Imaging
Spectrograph (STIS) onboard the \textit{Hubble Space Telescope} (\HST).

\footnotetext{Strictly speaking, \DZa\ is classed as a DZA due to the presence
of H$\alpha$ and H$\beta$ in its spectrum. However, for brevity we consider
this object as a DZ when collectively referring to all three stars.}

\section{Observations}
\label{sec:obs}

All three stars, \DZa, \DZb, and \DZc, have ample photometry spanning the
near-ultraviolet to the near-infrared. catalogues. Furthermore, as of
\emph{Gaia} eDR3, all three objects have 5-parameter astrometric solutions. The
available astrometric and photometric data for the three stars is given in
Table~\ref{tab:mags}.

\begin{table*}
    \centering
    \caption{\label{tab:mags}
    Astrometry and photometry for the white dwarfs observed with \HST\ STIS.
    For \emph{GALEX} NUV photometry, the first number is the value obtained by
    \emph{GALEX}, whereas the second number was calculated by integrating our
    STIS spectra over the \emph{GALEX} bandpass.
    }
     \begin{tabular}{lccc}
  \hline
   & \DZa & \DZb & \DZc \\
  \hline
  Ra                         &    09:56:45.147    &    10:38:09.091    &    15:35:05.810    \\
  Dec                        & $+$59:12:44.64     & $-$00:36:22.15     & $+$12:47:45.20     \\
  $\mu_\mathrm{Ra}$ [\masy]  & $-110.588\pm0.088$ & $-92.833\pm0.096$  & $-165.597\pm0.034$ \\
  $\mu_\mathrm{Dec}$ [\masy] & $ -31.488\pm0.131$ & $19.157\pm0.112$   & $-179.697\pm0.033$ \\
  parallax [mas]             & $7.4116\pm0.1219$  & $16.8686\pm0.0895$ & $51.9461\pm0.0344$ \\
  \hline
  \emph{GALEX} NUV     & $21.42\pm0.27/21.03\pm0.01$ & $19.67\pm0.08/19.74\pm0.01$ & $22.41\pm0.39/22.28\pm0.03$ \\ 
  SDSS $u$             & $18.950\pm0.021$            & $17.722\pm0.010$            & $18.039\pm0.013$            \\
  SDSS $g$             & $18.378\pm0.007$            & $16.993\pm0.004$            & $15.978\pm0.003$            \\
  SDSS $r$             & $18.403\pm0.009$            & $16.948\pm0.005$            & $15.469\pm0.003$            \\
  SDSS $i$             & $18.547\pm0.011$            & $17.079\pm0.005$            & $15.430\pm0.003$            \\
  SDSS $z$             & $18.747\pm0.035$            & $17.308\pm0.013$            & $15.497\pm0.005$            \\
  PanSTARRS $g$        & $18.3948\pm0.0067$          & $16.9718\pm0.0084$          & $15.9270\pm0.0057$          \\
  PanSTARRS $r$        & $18.4147\pm0.0057$          & $16.9568\pm0.0020$          & $15.5091\pm0.0005$          \\
  PanSTARRS $i$        & $18.5713\pm0.0046$          & $17.1095\pm0.0038$          & $15.4663\pm0.0042$          \\
  PanSTARRS $z$        & $18.7814\pm0.0149$          & $17.3335\pm0.0041$          & $15.5434\pm0.0070$          \\
  PanSTARRS $y$        & $18.8534\pm0.0216$          & $16.9401\pm0.0133$          & $15.5532\pm0.0072$          \\
  Gaia $G$             & $18.3839\pm0.0014$          & $16.9480\pm0.0012$          & $15.5905\pm0.0008$          \\
  Gaia $G_\mathrm{BP}$ & $18.4877\pm0.0182$          & $17.0644\pm0.0071$          & $15.9241\pm0.0061$          \\
  Gaia $G_\mathrm{RP}$ & $18.2951\pm0.0215$          & $16.7969\pm0.0125$          & $15.1300\pm0.0024$          \\
  2MASS $J$            & -                           & $16.836\pm0.142$            & $14.939\pm0.037$            \\
  2MASS $H$            & -                           & -                           & $14.728\pm0.045$            \\
  2MASS $K$            & -                           & -                           & $14.706\pm0.080$            \\
  UKIDSS $Y$           & -                           & $16.917\pm0.010$            & -                           \\
  UKIDSS $J$           & -                           & $16.848\pm0.014$            & -                           \\
  UKIDSS $H$           & -                           & $16.864\pm0.045$            & -                           \\
  UKIDSS $K$           & -                           & $16.661\pm0.052$            & -                           \\
  WISE $W1$            & $18.324\pm0.124$            & $16.756\pm0.047$            & $14.569\pm0.015$            \\
  WISE $W2$            & -                           & $16.788\pm0.128$            & $14.598\pm0.021$            \\
  \hline
\end{tabular}
\end{table*}

\subsection{Optical spectra}
\label{sec:obs_opt}

For \DZb\ and \DZc, in 2013 we obtained spectra covering 3100--10\,000\,\AA\
using the Intermediate-dispersion Spectrograph and Imaging System (ISIS) on the
William Herschel telescope (WHT). Details of the instrumental setup and basic
calibration have been summarised by \citet{hollandsetal17-1}. However we apply
an improved flux calibration to these spectra in order to increase the accuracy
of our spectroscopic fitting. The approach used is similar to the method for
correcting the flux calibration of SDSS BOSS spectra which we described in
\citet{hollandsetal17-1}. For the ISIS blue arm, we fit the differences between
the SDSS $u$ and $g$ magnitudes and the synthetic magnitudes of the WHT
spectra. This provided a wavelength dependent correction to the flux
calibration. Similarly for the ISIS red arm, we calculate synthetic magnitudes
for the $r$, $i$, and $z$ filters, and again fit the differences from the SDSS
photometry with a straight line to correct the flux calibration.

For \DZa\ we have obtained additional spectra using the OSIRIS spectrograph on
the GTC in service mode as part of ITP programme GTC1-16ITP. Four $u$-band
spectra using the R2500U grating were taken on the 4th January 2017, each with
900\,s exposure times. Additionally four $i$-band spectra were taken using the
R2500I grating on the 5th February 2017, again using 900\,s exposures. Standard
GTC service-mode calibration observations were also taken, i.e. bias frames
flat fields, arcs, and flux standards. Extraction of the 1D spectra was
performed using the \texttt{starlink} suite of software, with wavelength and
flux calibrations performed using \texttt{molly}\footnote{\texttt{molly}
software can be found at
\url{https://deneb.astro.warwick.ac.uk/phsaap/software/}}.

\subsection{\HST\ data}

The three white dwarfs \DZa, \DZb, and \DZc, were observed using \HST\ STIS,
using the G230L grating and the two arcsec wide slit from April to August 2016.
This setup covered wavelengths from 1580--3140\,\AA\ with a resolving power of
$R \simeq 500$. For \DZb\ and \DZc, these wavelengths effectively cover their
full SEDs when combined with our WHT spectra which go as blue as 3100\,\AA. For
\DZa, there remains a gap in observations between the \HST\ and GTC spectra,
though we found that all elements expected to be observed in that range already
had at least some lines at other wavelengths covered by our data. Standard
wavelength and flux calibrations were used throughout. The observations were
performed in time-tag mode, though we did not detect any signature of
variability in our data. Per-orbit spectra were coadded for the spectra used
for fitting, however for \DZc, guidance was lost during the final 240\,s of the
orbit, as evidenced by the time-tagged lightcurve. Therefore for the final
co-add, the fluxes of this sub-spectrum were scaled accordingly with the amount
of time lost. A summary of observations is provided in Table\,\ref{tab:hst}.
Table\,\ref{tab:mags} shows that synthetic \emph{GALEX} NUV photometry
calculated from the STIS data are consistent with the values measured by
\emph{GALEX} itself.

\begin{table}
  \centering
    \caption{\label{tab:hst}
    Observing log for \HST\ spectra for the three DZs. $N_\mathrm{orbit}$ is
    the number of orbits per object, and $t_\mathrm{exp}$ is the combined
    integration time for all orbits.
    }
    \begin{tabular}{lccc}
        \hline
         SDSS\,J & Obs. Date & $N_\mathrm{orbit}$ & $t_\mathrm{exp}$ \\
        \hline
         0956$+$5912 & 2016-04-20 & 2 &    5684 \\
         1038$-$0036 & 2016-05-15 & 2 &    5225 \\
         1535$+$1247 & 2016-08-11 & 4 & 10\,608 \\
         \hline
    \end{tabular}
\end{table}

\section{DZ models and spectral analysis}
\label{sec:atm}

The three white dwarfs were refitted using Koester (\citeyear{koester10-1})
model atmospheres, though with improvements to the code compared with our
previous analyses of these stars. In \citet{hollandsetal17-1}, we presented new
calculations for unified line profiles for the \Ion{Ca}{ii} H+K doublet,
\Ion{Mg}{ii} h+k doublet, and \Ion{Mg}{i} b triplet, though the latter was not
used in our previous analysis due to concerns over the atmospheric conditions
for this triplet potentially exceeding the validity of the unified theory,
instead using the more approximate quasi-static profiles
\citep{walkupetal84-1}.

Since then similar calculations for the Mg triplet have been used
\citep{blouin20-1} to show that the quasi-static profiles underestimate the
atmospheric abundances, potentially solving an Mg under-abundance issue
identified by in the \citet{hollandsetal17-1} sample by
\citet{turner+wyatt20-1}. We therefore also use our most up-to-date version of
the Mg triplet unified-profiles in this work. Furthermore, our models now
include a Ca-He unified profile for the \Ion{Ca}{i} 4227\,\AA\ resonance line,
first used by \citet{hollandsetal21-1}.

We also included photo-ionisation cross-sections (from TOPbase;
\citealt{cuntoetal93-1}) for dominating metals into both the atmospheric
structure calculations and spectral synthesis. We initially experimented to see
which cross sections were most important, so that those making effectively no
detectable difference were excluded during the fittings procedure. For all
three DZs, we used cross-sections for \Ion{Na}{i}, \Ion{Mg}{i}, \Ion{Al}{i},
\Ion{Si}{i}, \Ion{Ca}{i--ii}, and \Ion{Fe}{i}, though for \DZa\ \Ion{C}{i} was
also necessary.

Compared to the optical, the NUV contains a higher density of metal absorption
lines, many of which are broad and near saturated. Therefore an an increased
number of wavelength points were used in the atmospheric structure calculation,
in order to correctly sample the opacity contribution from all strong lines.

The models were fitted directly to the flux-calibrated spectra using a
$\chi^2$-minimization routine, with \Teff, white dwarf radius, and abundances
of all detected elements as free parameters at each step recalculating the
model structure and synthetic spectrum for self-consistent results. Previously
measured atmospheric parameters were initialised using the solutions from
\citet{hollandsetal17-1}. For newly identified elements (where lines were all
weak lines), initial abundances were adjusted manually until reasonable visual
agreement was found. For each object all available spectra were used in the
fit, with the models convolved to the instrumental resolution appropriate to
each spectrum, and with their individual $\chi^2$ summed to provide the final
$\chi^2$ for minimization. The radii were constrained via the \emph{Gaia}
parallaxes, and converted to surface gravities for input to the model
atmospheres using a mass-radius relation appropriate for thin hydrogen-layers
\citep{bedardetal20-1}. For \DZb\ and \DZc\, where hydrogen lines are not
observed, we adopted the approach of \citet{coutuetal19-1}, where hydrogen
abundances were fixed to the detection-limit. This approach has been found to
result in DZ masses more consistent with the mass distribution found for other
white dwarf spectral types \citep{coutuetal19-1}. Since these stars are within
150\,pc of the Sun, interstellar reddening $E(B-V)<0.01$ in all cases
\citep{greenetal18-1,greenetal19-1}, and so was assumed to be negligible.

\begin{table*}
    \centering
    \caption{\label{tab:params}
    Atmospheric parameters derived in this work (H2021), compared with previous
    results from \citet{hollandsetal17-1} (H2017). Abundances are logarithmic
    number abundances relative to helium, e.g. Ca corresponds to $\logX{Ca}$,
    and where their reported uncertainties do not include correlations between
    parameters.
    }
    \begin{tabular}{lcclcclcc}
\hline
SDSS\,J     & \multicolumn{2}{c}{0956$+$5912} & & \multicolumn{2}{c}{1038$-$0036} & & \multicolumn{2}{c}{1535$+$1247} \\
\hline                                                                                                                       
           & H2021 & H2017 &   & H2021 & H2017 &   & H2021 & H2017  \\
\hline                                                                                                                       
\Teff\ [K]  & $8100\pm100$            & $8800$  & & $7560\pm150$   & $7700$     & &  $5680\pm50$    & $5773$   \\  
$\log g$    & $8.02\pm0.04$           & $8.00$  & & $8.06\pm0.04$  & $8.00$     & &  $8.06\pm0.03$  & $8.00$   \\     
$M$ [\Msun] & $0.59\pm0.02$           & -       & & $0.61\pm0.03$  & -          & &  $0.61\pm0.02$  & -        \\     
H           & $-3.2\pm0.1$            & $-3.6$  & & $\leq-5.0$     & $\leq-4.8$ & &  $\leq -2.5$    & $-3$     \\  
C$^a$       & $[-5.2,-4.1]$           & -       & & -              & -          & &  -              & -        \\  
O$^b$       & $-4.1\pm0.2/-4.6\pm0.1$ & -       & & -              & -          & &  -              & -        \\  
Na          & $-6.8\pm0.2$            & $-6.60$ & & $-8.7\pm0.1$   & $-8.60$    & &  $-8.85\pm0.05$ & $-8.72$  \\  
Mg          & $-5.5\pm0.1$            & $-5.30$ & & $-6.6\pm0.1$   & $-6.80$    & &  $-6.9\pm0.1$   & $-7.36$  \\  
Al          & $-6.5\pm0.1$            & -       & & -              & -          & &  $-8.4\pm0.4$   & -        \\  
Si          & $-5.7\pm0.2$            & -       & & $-6.4\pm0.2$   & -          & &  $-6.5\pm0.3$   & -        \\  
Ca          & $-7.30\pm0.05$          & $-7.15$ & & $-7.60\pm0.05$ & $-7.85$    & &  $-8.50\pm0.05$ & $-8.61$  \\  
Ti          & $-8.9\pm0.2$            & -       & & $-9.5\pm0.1$   & $-9.60$    & & $-10.2\pm0.3$   & $-9.62$  \\  
Cr$^c$      & -                       & $-7.50$ & & -              & -          & &  $-9.6\pm0.3$   & $-9.25$  \\  
Mn          & $-9.2\pm0.2$            & -       & & $-9.2\pm0.2$   & -          & & $-10.0\pm0.1$   & -        \\  
Fe          & $-6.9\pm0.1$            & $-6.14$ & & $-7.4\pm0.1$   & $-7.40$    & &  $-7.6\pm0.1$   & $-7.57$  \\  
Ni          & -                       & -       & & $-8.6\pm0.1$   & -          & &  $-8.6\pm0.2$   & $-8.90$  \\  
\hline
\end{tabular}

  \newline
  Notes:~(a)~The carbon abundance is given as a lower/upper-limit pair based on
  the flux-suppression in the NUV and absence of molecular features in the
  optical. (b)~The oxygen abundance pair corresponds to the SDSS/GTC spectra
  respectively. (c)~A noise spike in the SDSS spectrum of \DZa\ was previously
  misidentified as the 5207\,\AA\ \Ion{Cr}{i} triplet, however the absence of
  other strong \Ion{Cr}{i--ii} lines rules out the detection of this element.
\end{table*}

The scaling of the models to the spectroscopic data in the continuum was used
to determine white dwarf angular sizes and their uncertainties since these are
anti-correlated with \Teff. Combining with the parallaxes and their errors
allowed us to measure the white dwarf radii and its covariance with \Teff.
Finally, we used a white dwarf mass-radius relation \citep{bedardetal20-1} to
determine the surface gravities and masses as well as their uncertainties.

To estimate parameter uncertainties, we calculated one-dimensional model grids
around the derived solution, in steps of 50\,K for \Teff, and steps of 0.1\,dex
for abundances. These models were then visually compared against all
spectroscopic data to estimate the level of uncertainty for each parameter. A
caveat of this approach, is that it does not consider covariance between
parameters, and so will lead to underestimates for some abundances, notably
between \Teff, H, Mg, Si, Ca, and Fe, which are typically mutually correlated.
Inspection of the parameter correlation matrices (derived from the covariance
matrices of our fits) indicates that this is at most a factor few\footnotemark.
Specifically Ca abundance uncertainties should be scaled by about a factor
four, whereas a factor two is more appropriate for H, Mg, Al, Si, Mn, Fe, and
Ni; or in other words, for most elements uncertainties including covariances
are at least 0.2\,dex. The remaining elements we have fitted (C, O, Na, Ti, and
Cr) only change the emergent flux where a limited set of narrow lines are
present, and thus are not correlated with other parameters.

\footnotetext{Scaling factors to correct the 1D error estimates can be found by
taking the square roots of the diagonal elements of the inverse correlation
matrices.}

\begin{figure*}
  \centering
  \includegraphics[width=\textwidth]{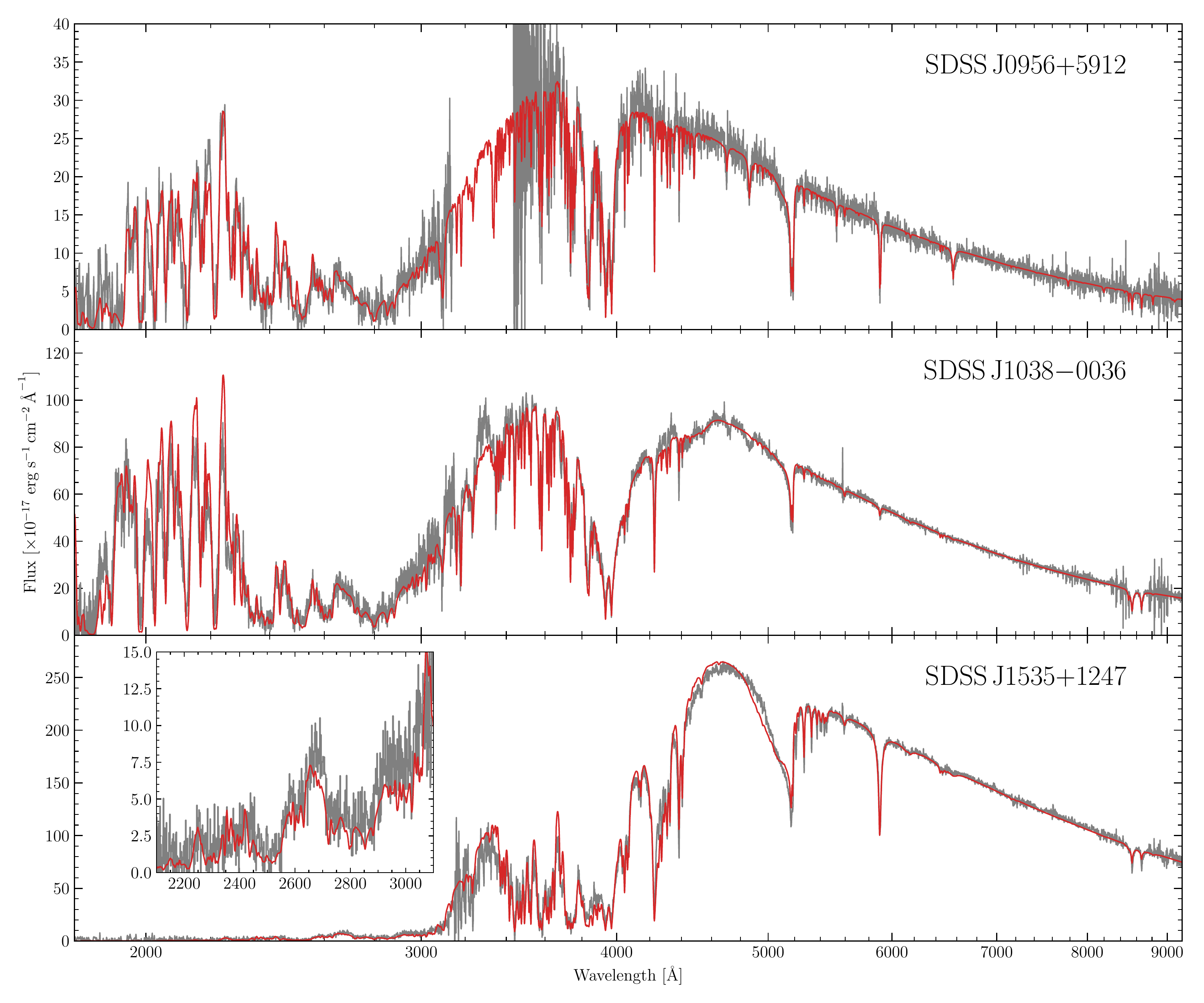}
  \caption{\label{fig:3spectra}
    Spectral energy distributions of all three white dwarfs covering both the
    NUV and optical with their best fitting models (red). For \DZc\ the insert
    shows the diminished flux in the NUV.
    }
\end{figure*}

The results to our fits are listed in Table\,\ref{tab:params}, with the best
fitting models shown against all spectra in Figure~\ref{fig:3spectra}. In all
three panels, the models adequately reproduce the spectral energy
distributions, both in the optical and the NUV. Nevertheless, a few
discrepancies remain. In particular, some missing absorption is observed in the
spectra of \DZa\ and \DZb\, in a narrow wavelength range around $\approx
2013$\,\AA. Searching with the NIST atomic spectra line database form
\citep{kramidaetal16-1}, we did not find any convincing candidates for missing
absorption lines. The \Ion{Co}{ii} resonance line at 2012.2\,\AA\ seemed the
most likely choice, but would require other strong \Ion{Co}{ii} lines to be
visible if this ion was present in the atmospheres of \DZa\ and \DZb. Other
minor discrepancies are visible at NUV wavelengths between regions of strong
line absorption (particularly for \DZb). However, because these differences
only affect a tiny fraction of the emergent flux, their influence on our
derived atmospheric parameters can be considered to be negligible.

\section{Effective temperatures}
\label{sec:Teff}

Our primary objective in this study was to assess the accuracy of the
atmospheric parameters derived from \citet{hollandsetal17-1}, when including
the NUV range in addition to only using the optical. For \DZb\ and \DZc, our
\Teff\ and abundance measurements were found to be consistent with the results
of \citet{hollandsetal17-1}, whereas for \DZa\ we found a substantially lower
temperature once the NUV was included (Table~\ref{tab:params}). Indeed our
earlier model from \citet{hollandsetal17-1} shows the NUV flux level to be
about 2--3 times higher than the flux observed with STIS. This poses the
question as to what is special about \DZa\ in comparison to \DZb\ and \DZc\
that such a large difference was found. We suggest some potential sources for
this systematic uncertainty affecting \DZa, though we do not find a conclusive
explanation.

All three objects were observed with the original SDSS spectrograph, with a
wavelength range extending as blue as 3800\,\AA. However, in the
\citet{hollandsetal17-1} analysis of \DZb\ and \DZc\ we had obtained additional
WHT spectra covering wavelengths down to 3100\,\AA. In all three spectra
(Figure~\ref{fig:3spectra}), a significant amount of flux is observed between
3000--3900\,\AA\ -- essentially the range between the extremely strong
\ion{Mg}{ii} h+k and \ion{Ca}{ii} H+K doublets. Therefore in the analysis of
\citet{hollandsetal17-1} this wavelength interval was already well constrained
for \DZb\ and \DZc, but practically unconstrained for \DZa\ (this wavelength
range is now largely covered by our GTC $u$-band spectrum). Even so, we only
found small differences between the flux levels in the $u$-band between the
older and newer models (where the main difference is instead the NUV as
described above). Though, we note that compared to our model from
\citet{hollandsetal17-1}, the wings of the Ca H+K resonance lines are in better
agreement with the optical spectra, and the \Ion{Mg}{ii} 4481\,\AA\ line is
also better matched in strength.

One difference which primarily affects \DZa\ between our model from
\citet{hollandsetal17-1} and our new results, is the inclusion of new elements
that were not previously detected. In particular Al is identified from a
\Ion{Al}{i} resonance doublet that sits between the Ca H+K lines. While the
doublet itself makes only a minor difference to the emergent flux of the
spectrum, the continuous opacity contributed by this element's photoionization
cross-section is considerable between between 1950--2250\,\AA, reducing the
emergent flux by half (compared to when the continuous opacity is excluded).
However, this can not explain the difference in \Teff\ we have found as
excluding this opacity source would necessitate an even lower \Teff\ in order
to remain broadly consistent with the observed fluxes.

Other recent analyses using similarly modern white dwarf model atmosphere codes
find virtually the same solution as the \citet{hollandsetal17-1} results when
using only the SDSS spectrum. Making use of the then newly available
\emph{Gaia} DR2 parallaxes, \citet{coutuetal19-1}, fitted the spectrum of \DZa,
assuming bulk Earth abundance ratios between metals, finding
$\Teff=8723\pm105$\,K, and a Ca abundance of $\logX{Ca}=-7.12\pm0.09$. More
recently \citet{blouin20-1} revisited the \citet{hollandsetal17-1} DZ sample
using their model atmosphere code. They elected to pre-compute a 6-dimensional
grid of models (\Teff, $\log g$; H, Ca, Mg, and Fe abundances) to use in their
analysis, and found $\Teff=8843\pm180$\,K, again highly similar to our previous
results. While these analyses find similar results to our previous solution
using optical spectra, we naturally cannot assess how these different models
would perform against the NUV data presented here. It is also important to note
that the models used by \citet{coutuetal19-1} and \citet{blouin20-1}, will
contain some differences in constituent physics such as the equation of state,
ionisation equilibrium, and unified line profiles used for the strongest metal
lines \citep{blouinetal18-1}, any of which could plausibly lead to differences
in results when analysing stars with extreme abundances, as is the case for
\DZa. However, direct comparisons between these model atmosphere codes is
beyond the scope of this work.

\section{Carbon detection in \texorpdfstring{\DZa}{SDSSJ0956+5912}}
\label{sec:carbon}

\begin{figure}
  \centering
  \includegraphics[angle=0,width=\columnwidth]{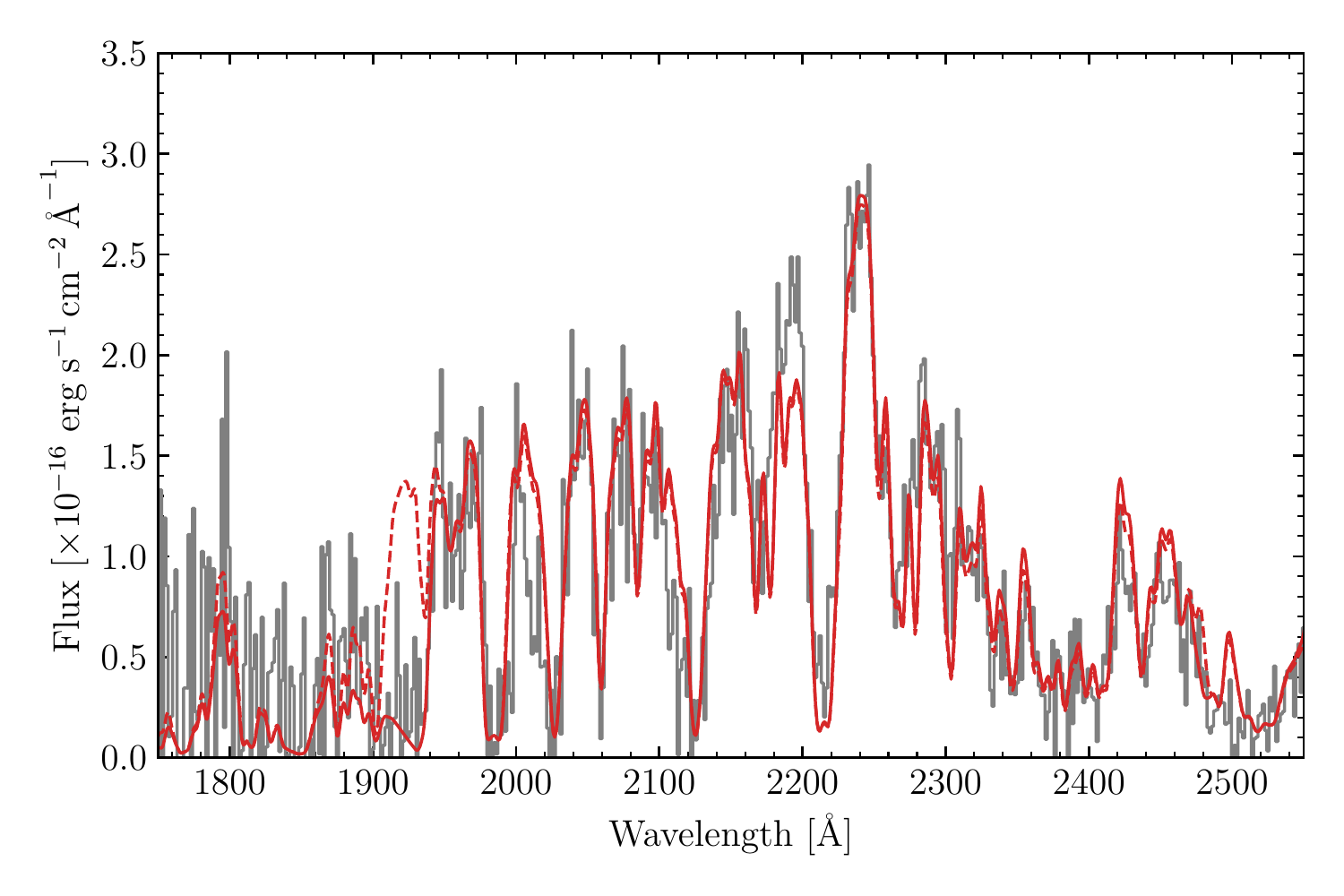}
  \caption{\label{fig:0956nuv}
  NUV spectrum of \DZa\ (grey). The best fitting model and a similar model with
  carbon removed are shown by the red solid/dashed curves, respectively.
    }
\end{figure}

For \DZa\ we report a tentative detection of photospheric carbon based on the
1931\,\AA\ line in the \HST\ spectrum. In Fig.~\ref{fig:0956nuv}, we show the
NUV spectrum of \DZa\ alongside two models. The solid red curve corresponds to
a model with $\logX{C}=-4.5$\,dex, whereas the dotted curve corresponds to a
model without any carbon. Evidently the carbon-rich model provides a better fit
in the range 1910--1940\,\AA\ where more flux would be expected in the absence
of carbon. Transitions from other elements were not found to be plausible, as
the only other strong lines (excluding trans-iron elements) in this range
belong to elements already measured at other wavelengths (Ti, Mn, Fe), but at
far too small abundances to cause strong absorption in the 1910--1940\,\AA\
range. Indeed the intrinsically strong lines from these other elements were
already included in our line list for both models in Figure\,\ref{fig:0956nuv}.

Detection of both carbon and heavy metals in white dwarf envelopes is rare,
though not unprecedented. Examples include Procyon B which was classified as a
DQ from its optical spectrum, though was reappraised as a DQZ following the
identification of magnesium from the \Ion{Mg}{ii} resonance lines in its UV
spectrum \citep{provencaletal02-1}. Three further DQZ stars have been
identified more recently by \citet{coutuetal19-1}. For objects classed DZQ,
i.e. where metal lines are dominant and carbon features are secondary, we have
only been able to identify one object in the literature: L\,119-34
(WD\,2216$-$657) as identified by \citet{wolffetal02-1}. Similarly to \DZa\
L\,119-34 has an optical spectrum dominated by metal lines
\citep{swanetal19-1}, but with the 1931\,\AA\ \Ion{C}{i} line also present in
the UV as demonstrated by its IUE spectrum.

However we acknowledge there are some reasons to treat this detection as
tentative. Firstly, this is the only carbon feature observed, without
additional corroborating evidence across the spectrum. Secondly, several
regions in the NUV are poorly modelled, and so it is feasible that the
1910--1940\,\AA\ range simply represents one more such region.

\begin{figure*}
  \centering
  \includegraphics[angle=0,width=\textwidth]{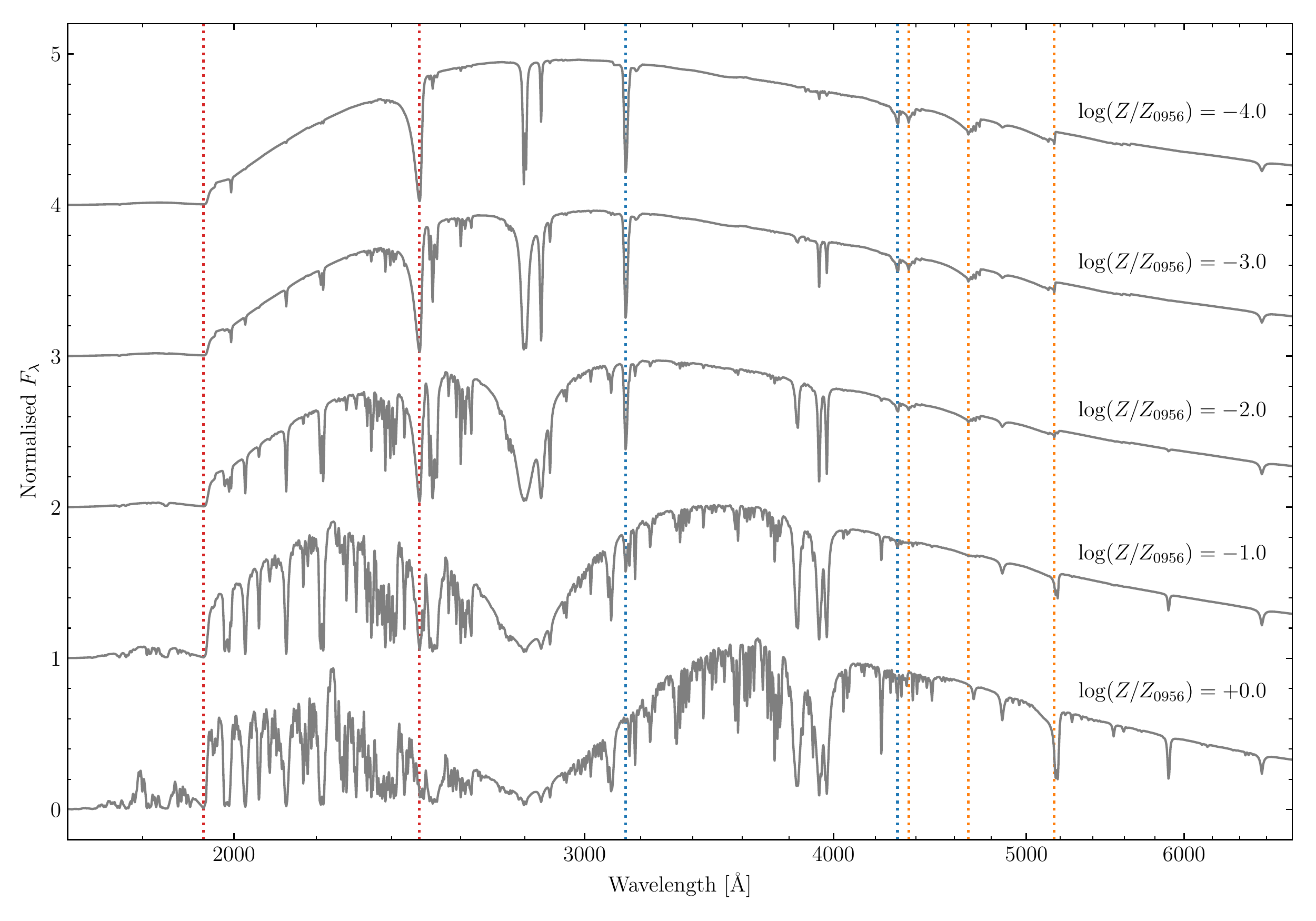}
  \caption{\label{fig:Cgrid}
  Grid of models starting at the measured abundances of \DZa, removing each
  metal (excluding carbon) in 1\,dex steps (bottom to top). As metals are
  removed, molecular carbon features become increasingly visible. The carbon
  abundance is set to $\logX{C}=-4.5$ in all cases. Wavelengths for atomic
  carbon, CH molecules, and C$_2$ molecules are indicated by the red, blue, and
  orange dotted lines, respectively.
    }
\end{figure*}

Assuming carbon is indeed present in the atmosphere, this element could have
either been accreted alongside the metals accreted by \DZa, or could be present
due to dredge up of carbon from the core, i.e. \DZa\ is a former DQ white dwarf
since converted to a DZQ. The first scenario can be considered unlikely as the
implied quantity of carbon is so large, that it dwarfs all other metals except
for oxygen (20--40 percent of the total metal mass). This allows for a roughly
one-to-one carbon-to-oxygen ratio -- inconsistent with other metal-rich white
dwarfs where carbon abundances have been explored \citep{wilsonetal16-1}. In
the dredge up scenario, the carbon abundance we found is certainly consistent
with other DQs with the same \Teff\
\citep{koester+kepler19-1,coutuetal19-1,blouinetal19-4}. However the DQs within
the published literature are predominantly identified via molecular features
from C$_2$ in the optical, yet these features are not seen for \DZa. Thus we
considered the possibility that the presence of metals could reduce the
observed strength of the Swan bands, preventing the detection of carbon in DZs.
We therefore constructed a small grid of models as a function of metal
abundance, starting with our best fit model to the SED of \DZa\ and
successively reducing metal abundances in 1\,dex steps, down to
$\log(Z/Z_{0956}) = -4$ (where $Z$ is the abundance of any metal and with the
carbon abundance held constant at $\logX{C} = -4.5$, and $Z_{0956}$ is the
corresponding abundance found for \DZa). The full grid of models are shown in
Figure\,\ref{fig:Cgrid}.

As hypothesised above, as the metal abundances are reduced, the Swan bands
become visible reaching their full strength at $\log(Z/Z_{0956}) \approx -3$.
In addition, other carbon features become visible in different parts of the
spectrum. The asymmetric 2479\,\AA\ \Ion{C}{i} line rapidly becomes apparent as
metals are reduced allowing it to be distinguished from the forest of metal
lines. At around 3145\,\AA\ the CH molecular band also becomes visible,
appearing weakly for $\log(Z/Z_{0956}) \approx -1$, and then much more strongly
as metal abundances are reduced further. An additional CH feature is seen at
4310\,\AA, but it is much weaker, and is instead comparable in strength to the
Swan bands.

We considered the possibility that the presence of metals provides more free
electrons to the atmosphere lowering the molecular number density at
equilibrium. However inspecting our models as a function of optical depth, we
found that the molecular abundances were not significantly changed. Instead we
found that at the high metal abundances of \DZa\, the opacity from metals
simply dominated the molecular opacity of carbon and so neither CH nor C$_2$
molecular bands could be detected.

DZ and DQ white dwarfs both correspond to at least $15\pm4$\,percent of the
local population with helium-rich atmospheres
\citep{mccleeryetal20-1}\footnote{Ratio based on confirmed helium-rich
atmospheres with $T_{\rm eff} > 5000$\,K and medium resolution spectroscopy in
the northern hemisphere 40\,pc sample.}. If the presence of carbon and metals
are independent, about 15\,percent of all DZ white dwarfs (corresponding to
2.25\,percent of white dwarfs with helium-rich atmospheres) would be expected
to show carbon lines, e.g. DZQ/DQZ spectral types. Large observed samples of DZ
stars contain a much lower fraction of DZQs/DQZs \citep{coutuetal19-1},
illustrating the dearth of these objects.

It is tempting to attribute the changing carbon feature strengths as a function
of metal abundance as an explanation for the apparent absence of known DZQs,
i.e. that the accretion of sufficiently large amounts of planetary material can
convert a DQ directly to a DZ (when only optical spectra are considered). While
this is a possibility for \DZa\ and other DZs with extreme metal-pollution,
this cannot explain the absence of DZQs/DQZs in a more general scenario.
Figure\,\ref{fig:Cgrid} shows that at $\log(Z/Z_{0956}) = -3$, when the C$_2$
bands reach their peak strength, the Ca H+K lines in the optical and Mg h+k
lines in the NUV still remain visible at a detectable level. Therefore a DQ
accreting small traces of rocky material would undoubtedly be identifiable as a
DQZ/DZQ, as was the case for Procyon B \citep[as mentioned
above,][]{provencaletal02-1}.

\section{The planetesimal accreted by \texorpdfstring{\DZa}{SDSSJ0956+5912}}
\label{sec:planet}

\subsection{Silicon and oxygen detections}

\DZa\ was previously reported as an outlier in terms of its large magnesium
abundance by \citet{hollandsetal18-1}. Assuming this magnesium to be part of
silicate rich material suggested silicon and oxygen should also be present in
similarly large quantities. For all three systems, the \HST\ data provided
coverage of detectable \Ion{Si}{i} absorption. The strongest groups of
(blended) lines are found near 1900\,\AA, 2060\,\AA, 1980\,\AA, 2129\,\AA,
2210\,\AA, and 2520\,\AA. However, for \DZa, we found that a further constraint
on the Si abundance was provided by the optical 3905\,\AA\ line which we had
previously missed.\footnotemark\ Re-inspection of all spectra included in
\citet{hollandsetal17-1} did not reveal the \Ion{Si}{i} 3905\,\AA\ line at any
of the other 230 white dwarfs in that work regardless of the spectral
signal-to-noise ratio, leaving only \DZa\ with this line present.

\footnotetext{While Si was not explicitly fitted in \citet{hollandsetal17-1},
lines from this element were still included in the modelling but with
abundances fixed relative to Ca. In hindsight, scaling Si with Mg would have
been a better choice.}

The measured silicon abundance is comparable to magnesium
(Table~\ref{tab:params}). Within rocky material, these elements are typically
part of the metal oxides MgO and SiO$_2$\footnotemark, thus suggesting around
three times as much oxygen present compared with magnesium. Unlike \Ion{Si}{i},
there are no strong \Ion{O}{i} lines in the NUV, only in the optical, where a
blend of three lines centred on 7775\,\AA\ (including the
$\sim80$\,km\,s$^{-1}$ redshift measured from other photospheric lines), ought
to be the strongest \Ion{O}{i} feature at this \Teff. Indeed, a weak, but clear
\Ion{O}{i} feature is detected at 7775\,\AA\ in the SDSS spectrum of \DZa,
corresponding to a measured abundance of $\logX{O} = -4.1\pm0.2$\,dex -- much
higher than can be expected for a combination of the oxides MgO and SiO$_2$.

\footnotetext{The base metal oxides MgO and SiO$_2$ needn't be present in this
form prior to accretion, and could instead by part of more complex minerals
such as MgSiO$_3$ (enstatite) or Mg$_2$SiO$_4$ (forsterite). Since we only
measure the individual atomic abundances, we refrain from further speculation
on the precise mineralogy of the accreted parent body and instead refer to the
base metal oxides.}

\begin{figure*}
  \centering
  \includegraphics[angle=00,width=\textwidth]{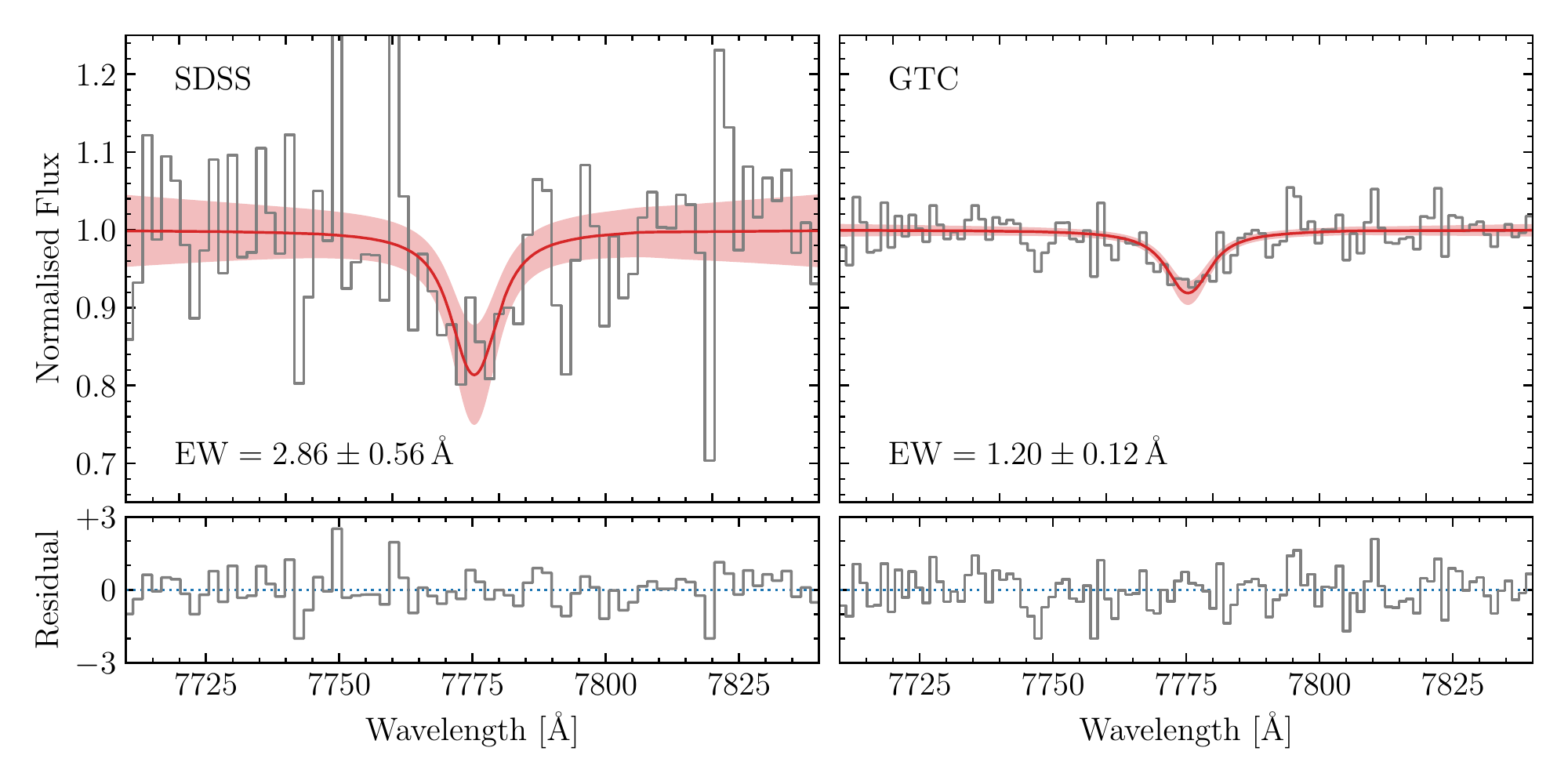}
  \caption{\label{fig:ew}
    Equivalent width of the \Ion{O}{i} blend for \DZa, as measured from the
    SDSS and GTC spectra. The red shaded regions indicate the 2$\sigma$
    confidence region around the best fit. The residuals between the data and
    fit are scaled by the data uncertainties.
    }
\end{figure*}

To better measure the oxygen abundance, we obtained a higher quality spectrum
with the OSIRIS instrument on the GTC, and using the $i$-band VPH grating
(Section\,\ref{sec:obs_opt}). The \Ion{O}{i} blend remains visible in the GTC
spectrum, though surprisingly with the line strength reduced compared to the
SDSS spectrum (Figure\,\ref{fig:ew}), and the corresponding abundance
diminished to $\logX{O}=-4.6\pm0.1$\,dex (Table~\ref{tab:params}). We do not
observe similar changes in line strength for other transitions covered by both
the SDSS and GTC spectra. In the GTC $i$-band spectrum no change is seen for
the \Ion{Ca}{ii} triplet, nor the 8807\,\AA\ \Ion{Mg}{i} line, compared with
the SDSS spectrum. Furthermore, the approximately twenty narrow absorption
lines observed in both the GTC $u$-band spectrum and SDSS spectrum are also
consistent in strength.

To precisely quantify the apparent change in oxygen line strength, we estimated
the equivalent widths in both spectra, adopting an approach based on the
optimal-extraction algorithm developed for CCD spectroscopy \citep{horne86-1}.
This method essentially performs a weighted sum across a row of pixels, where
the weights are chosen to maximise the sum signal to noise ratio. This has the
advantage of ignoring the contribution of variance far outside the
line-profile, as would be the case with a straight sum. One important
ingredient to this approach is that an accurate shape of the line-profile must
be known in advance, which fortunately is possible in our case. We extract this
shape directly from the model spectrum, convolving to the instrumental
resolution, and redshifting to the frame as measured from other spectral lines.

Indeed the equivalent widths were found to be in disagreement
(Figure\,\ref{fig:ew}), where the line strengths are seen to change by
$1.66\pm0.57$\,\AA, i.e. about $3\sigma$ difference. We are unable to offer a
clear explanation for why the oxygen triplet appears to have changed strength.
As previously mentioned, the moderate 80\,\kms\ velocity-shift of the oxygen
triplet matches that of the other metal lines in the spectrum confirming it is
of photospheric origin. While a weak magnetic field combined with rotation
could plausibly lead to line strength variability, we do not observe similar
variation between other lines. Furthermore, no evidence of Zeeman splitting is
observed in any lines, in particular the components of the \Ion{Ca}{ii}
infra-red triplet which is particularly sensitive to small fields due to its
red wavelength and large Land\'e-$g$ factors, allowing us to place an
upper-limit of 40\,kG from the GTC $i$-band spectrum. A heterogeneous abundance
distribution is also ruled out as the horizontal spreading timescale for metals
across the surface of cool white dwarfs with helium-dominated atmospheres
should be around four orders of magnitude faster than the sinking timescale
\citep{cunninghametal21-1}. With no ready explanation for the peculiar change
in abundances, we conduct the rest of our analysis into \DZa\ referring to
results for both the SDSS and GTC abundances.

While substantially lower than for the SDSS spectrum ($-4.1$\,dex), the GTC
abundance ($-4.6$\,dex) remains larger than expected for a composition
dominated by MgO and SiO$_2$. Given that the other metals in \DZa\ are
essentially traces compared with magnesium, silicon, and oxygen (for instance
iron is about 1\,dex depleted relative to its bulk Earth ratio compared with
magnesium), it is clear that this oxygen over-abundance cannot be entirely
attributed to other metal oxides. Past work has shown that oxygen-excesses can
be explained as water making up a large fraction of the accreted material
\citep{farihietal13-1, raddietal15-1}. In those reported cases, the white
dwarfs are of spectral type DBZA, where the trace hydrogen serves as an
additional signature of water accretion \citep{gentileetal17-1}, though unlike
metals hydrogen does not sink out of the white dwarf envelope, instead
accumulating over all accretion episodes. Like those previous cases, \DZa\ also
shows hydrogen traces, at an abundance of $\logX{H}=-3.2\pm0.1$\,dex. The water
accretion interpretation will be discussed in more detail in
Section~\ref{sec:water}.

\subsection{Envelope parameters of \texorpdfstring{\DZa}{SDSSJ0956+5912}}
\label{sec:env}

\begin{table*}
  \centering
    \caption{\label{tab:cvz}
    Envelope parameters for \DZa. The first and second sets of columns
    correspond to envelope models calculated with and without convective
    overshoot extending the size of the convection zone by one pressure scale
    height, respectively. $\log q$ is the logarithm (base-10) of the fractional
    convection zone mass, $M_\mathrm{cvz}/M_\mathrm{wd}$. Diffusion timescales
    for each element, $\tau_Z$ are in units of years; the convection zone
    masses for each element, $m_Z$, in g; and the diffusion fluxes out of the
    base of the convection zone, $\dot m_Z$, in g\,s$^{-1}$ (with all three
    quantities expressed as base-10 logarithms). Oxygen masses and diffusion
    fluxes correspond to the abundance as measured in the SDSS spectrum, where
    the corresponding values for the GTC spectrum can be obtained by
    subtracting 0.50\,dex. Diffusion timescales were calculated for nickel,
    although since this element is not detected, values of $m_Z$, and $\dot
    m_Z$ cannot be calculated. 
    }
    \begin{tabular}{lccccccc}
    \hline
             & \multicolumn{3}{c}{Overshoot} & & \multicolumn{3}{c}{No overshoot} \\
    $\log q$ & \multicolumn{3}{c}{$-5.224$} & & \multicolumn{3}{c}{$-5.658$} \\
    \hline
             & $\log\tau_Z$ & $\log m_Z$ & $\log \dot m_Z$ & & $\log\tau_Z$ & $\log m_Z$ & $\log \dot m_Z$ \\
    \hline
    H  & -       & $24.05$ & -       & & -       & $23.61$ & -       \\
    C  & $6.460$ & $23.82$ & $ 9.87$ & & $6.114$ & $23.39$ & $ 9.78$ \\
    O  & $6.375$ & $24.35$ & $10.48$ & & $6.026$ & $23.91$ & $10.39$ \\
    Na & $6.239$ & $21.80$ & $ 8.07$ & & $5.888$ & $21.37$ & $ 7.98$ \\
    Mg & $6.241$ & $23.13$ & $ 9.39$ & & $5.887$ & $22.69$ & $ 9.31$ \\
    Al & $6.195$ & $22.17$ & $ 8.48$ & & $5.842$ & $21.74$ & $ 8.40$ \\
    Si & $6.201$ & $22.99$ & $ 9.29$ & & $5.846$ & $22.56$ & $ 9.21$ \\
    Ca & $6.088$ & $21.55$ & $ 7.96$ & & $5.730$ & $21.11$ & $ 7.88$ \\
    Ti & $5.999$ & $20.02$ & $ 6.53$ & & $5.643$ & $19.59$ & $ 6.45$ \\
    Mn & $5.953$ & $19.78$ & $ 6.33$ & & $5.597$ & $19.35$ & $ 6.25$ \\
    Fe & $5.956$ & $22.09$ & $ 8.64$ & & $5.599$ & $21.66$ & $ 8.56$ \\
    Ni & $5.947$ & -       & -       & & $5.589$ & -       & -       \\
     \hline
\end{tabular}

\end{table*}

To fully analyse the composition of the accreted material requires knowledge of
the white dwarf envelope parameters, specifically the convection zone mass and
the diffusion timescales for each element in the atmosphere. The extreme metal
abundances (as well as the large quantity of trace hydrogen) have a strong
affect on the atmospheric structure, and in turn upon the envelope parameters.
It is therefore insufficient to use a tabulated grid of values as function of
\Teff\ and $\log g$, where trace metal abundances have little effect on the
structure, and instead bespoke calculations are required. We calculated these
convection zone parameters using the latest version of the Koester envelope
code \citep{koesteretal20-1}, using our best fitting model for \DZa\ as a
boundary condition at the top of the envelope. We calculated two envelope
models, firstly with the chemically mixed region determined purely from mixing
length theory, and secondly with convective-overshoot extending the chemically
mixed region by one pressure scale height (results presented in
Table~\ref{tab:cvz}). The effect of this overshoot acts to increase diffusion
timescales by $0.35$\,dex or a factor of $2.2$. Compared to models with only
minute traces of metals (but the same $\Teff$ and $\log g$), the bespoke models
result in smaller convection zones (by 0.67\,dex), and consequently also have
shorter diffusion timescales (by about 0.6\,dex).

\subsection{An oxygen excess}

\begin{figure}
    \centering
    \includegraphics[width=\columnwidth]{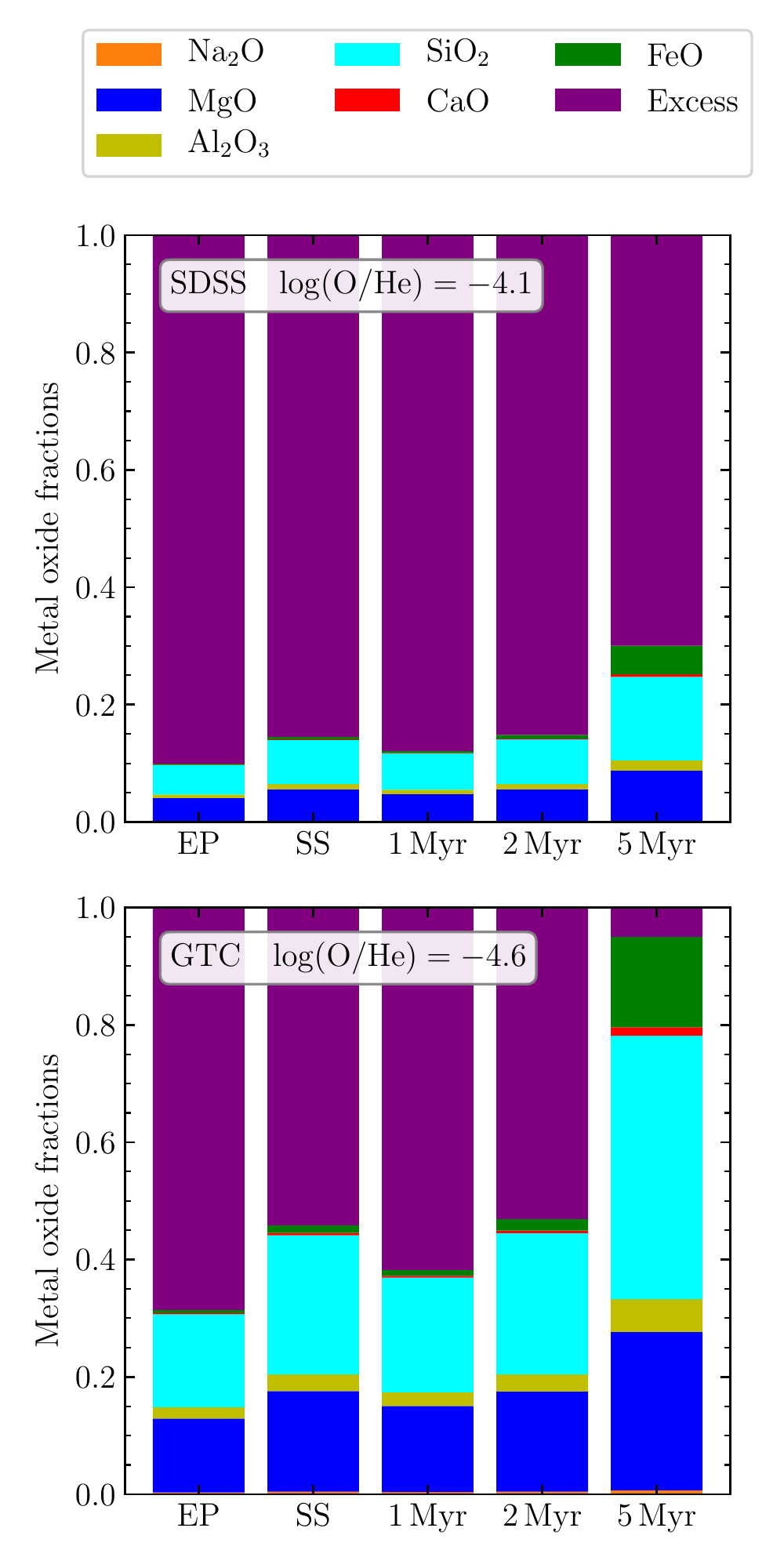}
    \caption{\label{fig:OXS}
    Metal oxide fractions and oxygen excess for the material accreted by \DZa\
    under different assumptions. The top and bottom panels reflect the
    different abundances measured for the SDSS and GTC spectra respectively.
    \textbf{EP}: Early phase accretion, where atmospheric abundances directly
    reflect the accreted composition. \textbf{SS}: In the steady-state phase
    accretion balances diffusion of metals out of the outer convection zone.
    \textbf{1--5\,Myr}: In the decreasing phase the observed a large fraction
    of the accreted metals may have already been depleted out of the outer
    convection zone, albeit at different rates. The three columns demonstrate
    the parent body composition assuming accretion occurring 1, 2 and 5\,Myr
    ago.
    }
\end{figure}

We apply the same analysis method as that used by \citet{farihietal13-1} to
quantify the level of excess oxygen. Specifically, we assume metals heavier
than oxygen\footnotemark\ were originally in metal oxide form, e.g. MgO,
Al$_2$O$_3$, and SiO$_2$. The corresponding amounts of oxygen are then
subtracted from the total oxygen budget revealing any remaining oxygen as an
excess. However, depending at which stage of accretion \DZa\ is observed at,
different treatments are needed. In the early phase of accretion, diffusion
processes have not had time to alter the accreted abundances, and so
atmospheric abundance ratios can be taken as reflective of the accreted parent
body. If diffusion rates are much shorter than the length of an accretion
episode, then a steady-state equilibrium may be reached where accretion
balances diffusion processes out of the atmosphere. In this case, the
abundances are scaled by dividing by the corresponding diffusion timescales.
Finally, we consider the decreasing phase, where a significant fraction of the
accreted material may have already sunk out of the outer convection zone. In
this case we re-scale each elemental abundance by a factor $\exp(t/\tau_Z)$
where $t$ is the time since accretion (using values of 1, 2, and 5\,Myr), and
$\tau_Z$ is the diffusion timescale for an element $Z$. Our results from this
analysis are shown in Figure\,\ref{fig:OXS} for oxygen abundances as measured
from both the SDSS and GTC spectra. Note that we do not consider titanium or
manganese oxides, as Ti and Mn abundances are more than an order of magnitude
lower than the other metals that we consider.

\footnotetext{We assume any carbon present to be unrelated to the accreted
abundances in our analysis, given that C/O ratios are typically small
\citep{wilsonetal16-1} and that the accreted material lacks other volatile
elements such as nitrogen and sulphur typical of cometary material
\citep{xuetal17-1}.}

\begin{figure}
    \centering
    \includegraphics[width=\columnwidth]{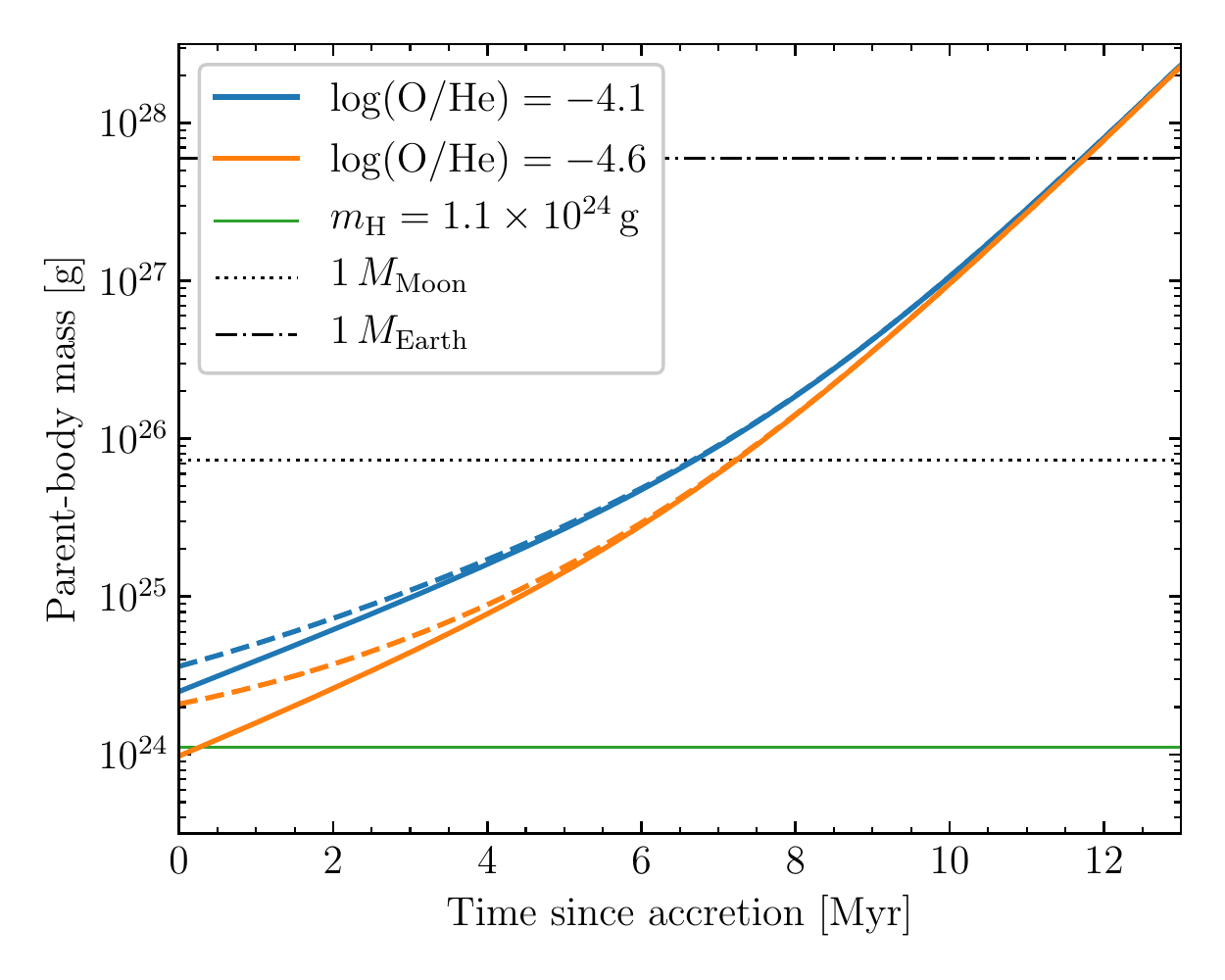}
    \caption{\label{fig:exo_mass}
    The inferred mass of the planetary object accreted by \DZa\ as a function
    of how long ago accretion occurred. This is determined separately for both
    the SDSS and GTC oxygen abundances (blue and orange curves respectively).
    The parent-body mass when hydrogen is included in the calculations is shown
    by the dashed lines, and the hydrogen mass in the convection zone,
    $m_\mathrm{H}$, is shown by the solid green line. The curves converge for
    accretion times beyond 10\,Myr as oxygen becomes a relatively minor
    component of the total parent-body mass. All calculations assume a
    convection zone size and diffusion timescales corresponding to one-pressure
    scale height of convective-overshoot.
    }
\end{figure}

It is evident from Figure\,\ref{fig:OXS} that in early phase accretion, i.e.
using the abundances directly as measured from the atmosphere, \DZa\ boasts the
largest oxygen excess observed for any white dwarf so far
\citep{farihietal13-1,raddietal15-1,hoskinetal20-1,izquierdoetal21-1}, where
the excess oxygen fraction is 90\,percent for the SDSS measurement, and 70
percent for the GTC spectrum. These excesses are reduced somewhat in the
steady-state phase, though such a scenario is unlikely for \DZa\ as diffusion
timescales in the range 0.9--2.4\,Myr are likely to exceed the length of any
accretion episode. Furthermore the absence of a flux excess in the
\textit{WISE} photometry suggests the absence of a disc. More significant
differences are seen for the decreasing phase, particularly at 5\,Myr. Here the
difference between the $\logX{O} = -4.1/-4.6$\,dex scenarios becomes vastly
more important. For the SDSS measurement, even at 5\,Myr the excess remains at
$\simeq 70$\,percent. By contrast, for the GTC measurement, if accretion
occurred 2\,Myr ago, a moderate excess of about 45 percent is seen (comparable
to the steady-state value for GD\,61 as found by \citealt{farihietal13-1}), and
as the time since accretion approaches 5\,Myr, the oxygen-excess is rapidly
erased, allowing for a parent body composed almost entirely of metal oxides
with little excess oxygen and a total mass of $10^{25}$\,g
(Figure~\ref{fig:exo_mass}). If iron was instead assumed to be present in the
form of solid metal then the oxygen fraction ($\simeq 15$\,percent at 5\,Myr)
previously ascribed to FeO will correspond to an excess instead (though this
excess would also be erased within an additional 1.2\,Myr).

\subsection{Evidence for accretion of water ice}
\label{sec:water}

In all previously analysed DBZs where oxygen-excesses have been identified,
moderate traces of hydrogen have also been found polluting the white dwarf
atmospheres
\citep{farihietal13-1,raddietal15-1,hoskinetal20-1,izquierdoetal21-1}. Since
the excess oxygen must have been accreted from a solid-component of the parent
body, an additional molecule is required, with water-ice being the obvious
candidate \citep{gentileetal17-1}. For \DZa\ a hydrogen abundance of
$\logX{H}=-3.2$ was measured, and is evidently more than twice the oxygen
abundance (i.e. as would be expected for H$_2$O) measured from the SDSS
spectrum ($\logX{O}=-4.1$). This difference can be resolved by two possible
explanations. If the metals at \DZa\ were accreted relatively recently (early
phase), then only a small fraction of the observed hydrogen can have been
delivered in that accretion episode. This is simply resolved if the remaining
hydrogen was delivered from prior accretion episodes in the history of \DZa,
where the metals have long since sunk out of the photosphere, but with the
integrated hydrogen mass continuing to remain in the white dwarf convection
zone. Alternatively, it is possible for all the photospheric hydrogen to have
been delivered with the metals if we assume \DZa\ is observed far along the
decreasing phase. Given the $1.1\times10^{24}$\,g of hydrogen in the white
dwarf convection-zone, which does not evolve over time
(Figure~\ref{fig:exo_mass}), this necessitates a water mass of $1.0\times
10^{25}$\,g within the accreted planetesimal. Specifically, assuming the SDSS
abundance, the oxygen-excess mass can be matched to the present-day hydrogen
abundance if accretion occurred $\simeq4$\,Myr ago, implying around 60\,percent
of the parent body mass to be composed of water-ice, i.e. a $1.7\times
10^{25}$\,g overall parent-body mass ($\sim 0.2$ lunar masses). For comparison,
the dwarf planet Ceres is estimated to have a water-ice mass-fraction between
17 and 25\,percent \citep{mccord+sotin05-1}. Of the remaining mass around
$5\times10^{24}$\,g would be composed of MgO and SiO$_2$, and around
$1\times10^{24}$\,g composed of FeO. 

The above argument cannot be made assuming the GTC oxygen abundance: while
increasing the look-back time increases the overall mass of accreted oxygen, as
previously discussed the oxygen-excess is erased at about 5\,Myr leaving no
oxygen that could be associated with the trace hydrogen (the absolute oxygen
excess peaks about one order of magnitude less than required to fully
correspond to the hydrogen abundance). However, if the planetesimal was
accreted more recently, e.g. 2\,Myr ago, a substantial oxygen excess remains
($\simeq50$ percent at 2\,Myr), still necessitating water-ice to be a major
component of the parent body. In this case however, only \emph{some} of the
hydrogen in the atmosphere of \DZa\ can be associated with the present metal
abundances, with the remaining hydrogen having been accumulated from previous
episodes of water-rich accretion.

\subsection{Accretion of core material as an unlikely scenario}

Recently \cite{harrisonetal21-1} reanalysed most of the
\citet{hollandsetal17-1} sample. While recognising that \DZa\ shows Mg-enhanced
abundances at present, their analysis led to the conclusion that this star
accreted core-like material of which most of the Fe and Ni has diffused out of
the atmosphere. We show that this scenario is unlikely. We first consider the
composition of accreted planetary cores which can be expected to only show
traces of Mg. In \citet{hollandsetal17-1,hollandsetal18-1}, the white dwarf
\sdss{0823}{+}{0546} is among the most-core like with abundances dominated by
Fe and Ni, and $\logXY{Mg}{Fe} = -1.15$. For \DZa, we instead find
$\logXY{Mg}{Fe} = +1.4$. From Table~\ref{tab:cvz} the diffusion timescale for
Mg is around twice that for Fe, and so given enough time the Mg/Fe ratio could
switch to Mg-dominated. Using equation~3 from \citet{hollandsetal18-1}, it can
be demonstrated that for Mg/Fe to increase by 2.55\,dex requires changes of
$\Delta\logX{Mg}=-2.75$\,dex and $\Delta\logX{Fe}=-5.30$\,dex over a period of
11\,Myr. Ignoring changes in the convection zone size due to the extreme
implied metal abundances, the parent-body itself would have a mass of $\simeq
3\times10^{27}$\,g, i.e. about half the mass of the Earth
(Figure\,\ref{fig:exo_mass}). Given that planetary systems contain only a few
objects with masses $>10^{27}$\,g we consider this scenario extremely unlikely
in comparison to accretion of a body of $\sim 10^{25}$\,g, for which far more
bodies may be available and can be more easily scattered towards the white
dwarf.

In summary, the material accreted by \DZa\ resembles water ice mixed with
magnesium silicates under the assumption of accretion in the last $4$\,Myr.
While a less extreme composition may be possible assuming accretion occurred
beyond $10$\,Myr ago, such a scenario is a-priori unlikely due to the required
parent-body mass exponentially increasing with time post-accretion, which is in
tension with a mass-distribution of exoplanetesimals rapidly decreasing towards
higher masses.

\section{Conclusions}
\label{sec:conc}

In general we find that while NUV spectroscopy provides additional constraints
on DZ white dwarf atmospheric parameters, optical spectra -- which can be
obtained from ground based observatories -- are sufficient for an accurate
analysis in most cases. However, in the case of \DZa, the addition of NUV
spectra resulting in a larger discrepancy, indicating that additional care
should be taken when modelling spectra of warm, highly metal-rich DZs when only
optical spectra are available. For \DZa\ we measured extreme abundances of
oxygen, magnesium, and silicon; indicating the accreted parent body was likely
composed of MgO, SiO$_2$, and water-ice, with a total parent-body mass of up to
$\simeq0.2$ lunar masses (for accretion having occurred in the last 4\,Myr).
Furthermore we tentatively detect carbon at \DZa\ suggesting it may have had a
spectral type DQ before accreting its present abundances, but we demonstrate
that this cannot explain the lack of observed of DZQs/DQZs more generally.

\section*{Acknowledgements}

The authors acknowledge the referee, P. Dufour, who provided useful comments
that improved the quality of this manuscript. This research received funding
from the European Research Council under the European Union’s Horizon 2020
research and innovation programme number 677706 (WD3D).  
BTG was supported by a Leverhulme Research Fellowship and the UK STFC grant
ST/T000406/1.
Based on observations made with the Gran Telescopio Canarias (GTC), installed
in the Spanish Observatorio del Roque de los Muchachos of the Instituto de
Astrof\'isica de Canarias (IAC), in the island of La Palma, and on observations
made with the William Herschel Telescope (WHT) which is operated on the island
of La Palma by the Isaac Newton Group of Telescopes in the Spanish Observatorio
del Roque de los Muchachos of the IAC, and on observations made with the
NASA/ESA Hubble Space Telescope, obtained at the Space Telescope Science
Institute, which is operated by the Association of Universities for Research in
Astronomy, Inc., under NASA contract NAS 5-26555. These observations are
associated with programmes \#14128. Data for this paper have been obtained
under the International Time Programme of the CCI (International Scientific
Committee of the Observatorios de Canarias of the IAC).
\section*{Data Availability}

All spectroscopic data used in this work were either obtained from public
databases (SDSS), or are now publicly available from the respective observatory
archives having exceeded their proprietary periods. For all \HST\ data, the
programme number is 14128. WHT ISIS spectra can be found from the CASU archive
on nights 20131228 and 20131229. GTC spectra can be found from the GTC archive
with programme number GTC1-16ITP.




\bibliographystyle{mnras}
\bibliography{aamnem99,aabib}

\begin{thebibliography}{}
\makeatletter
\relax
\def\mn@urlcharsother{\let\do\@makeother \do\$\do\&\do\#\do\^\do\_\do\%\do\~}
\def\mn@doi{\begingroup\mn@urlcharsother \@ifnextchar [ {\mn@doi@}
  {\mn@doi@[]}}
\def\mn@doi@[#1]#2{\def\@tempa{#1}\ifx\@tempa\@empty \href
  {http://dx.doi.org/#2} {doi:#2}\else \href {http://dx.doi.org/#2} {#1}\fi
  \endgroup}
\def\mn@eprint#1#2{\mn@eprint@#1:#2::\@nil}
\def\mn@eprint@arXiv#1{\href {http://arxiv.org/abs/#1} {{\tt arXiv:#1}}}
\def\mn@eprint@dblp#1{\href {http://dblp.uni-trier.de/rec/bibtex/#1.xml}
  {dblp:#1}}
\def\mn@eprint@#1:#2:#3:#4\@nil{\def\@tempa {#1}\def\@tempb {#2}\def\@tempc
  {#3}\ifx \@tempc \@empty \let \@tempc \@tempb \let \@tempb \@tempa \fi \ifx
  \@tempb \@empty \def\@tempb {arXiv}\fi \@ifundefined
  {mn@eprint@\@tempb}{\@tempb:\@tempc}{\expandafter \expandafter \csname
  mn@eprint@\@tempb\endcsname \expandafter{\@tempc}}}

\bibitem[\protect\citeauthoryear{{Allard}, {Leininger}, {Gad{\'e}a},
  {Brousseau-Couture}  \& {Dufour}}{{Allard} et~al.}{2016a}]{allardetal16-1}
{Allard} N.~F.,  {Leininger} T.,  {Gad{\'e}a} F.~X.,  {Brousseau-Couture} V.,
  {Dufour} P.,  2016a, \mn@doi [A\&A] {10.1051/0004-6361/201527826}, \href
  {http://adsabs.harvard.edu/abs/2016A%26A...588A.142A} {588, A142}

\bibitem[\protect\citeauthoryear{{Allard}, {Guillon}, {Alekseev}  \&
  {Kielkopf}}{{Allard} et~al.}{2016b}]{allardetal16-2}
{Allard} N.~F.,  {Guillon} G.,  {Alekseev} V.~A.,   {Kielkopf} J.~F.,  2016b,
  \mn@doi [A\&A] {10.1051/0004-6361/201628781}, \href
  {http://adsabs.harvard.edu/abs/2016A%26A...593A..13A} {593, A13}

\bibitem[\protect\citeauthoryear{{B{\'e}dard}, {Bergeron}, {Brassard}  \&
  {Fontaine}}{{B{\'e}dard} et~al.}{2020}]{bedardetal20-1}
{B{\'e}dard} A.,  {Bergeron} P.,  {Brassard} P.,   {Fontaine} G.,  2020, arXiv
  e-prints, \href {https://ui.adsabs.harvard.edu/abs/2020arXiv200807469B} {p.
  arXiv:2008.07469}

\bibitem[\protect\citeauthoryear{{Blouin}}{{Blouin}}{2020}]{blouin20-1}
{Blouin} S.,  2020, \mn@doi [MNRAS] {10.1093/mnras/staa1689}, \href
  {https://ui.adsabs.harvard.edu/abs/2020MNRAS.tmp.1829B} {}

\bibitem[\protect\citeauthoryear{{Blouin} \& {Dufour}}{{Blouin} \&
  {Dufour}}{2019}]{blouinetal19-4}
{Blouin} S.,  {Dufour} P.,  2019, \mn@doi [MNRAS] {10.1093/mnras/stz2915},
  \href {https://ui.adsabs.harvard.edu/abs/2019MNRAS.490.4166B} {490, 4166}

\bibitem[\protect\citeauthoryear{{Blouin}, {Dufour}  \& {Allard}}{{Blouin}
  et~al.}{2018}]{blouinetal18-1}
{Blouin} S.,  {Dufour} P.,   {Allard} N.~F.,  2018, \mn@doi [ApJ]
  {10.3847/1538-4357/aad4a9}, \href
  {https://ui.adsabs.harvard.edu/abs/2018ApJ...863..184B} {863, 184}

\bibitem[\protect\citeauthoryear{{Blouin}, {Dufour}, {Allard}, {Salim}, {Rich}
  \& {Koopmans}}{{Blouin} et~al.}{2019a}]{blouinetal19-1}
{Blouin} S.,  {Dufour} P.,  {Allard} N.~F.,  {Salim} S.,  {Rich} R.~M.,
  {Koopmans} L.~V.~E.,  2019a, \mn@doi [ApJ] {10.3847/1538-4357/ab0081}, \href
  {https://ui.adsabs.harvard.edu/abs/2019ApJ...872..188B} {872, 188}

\bibitem[\protect\citeauthoryear{{Blouin}, {Allard}, {Leininger}, {Gad{\'e}a}
  \& {Dufour}}{{Blouin} et~al.}{2019b}]{blouinetal19-2}
{Blouin} S.,  {Allard} N.~F.,  {Leininger} T.,  {Gad{\'e}a} F.~X.,   {Dufour}
  P.,  2019b, \mn@doi [ApJ] {10.3847/1538-4357/ab1266}, \href
  {https://ui.adsabs.harvard.edu/abs/2019ApJ...875..137B} {875, 137}

\bibitem[\protect\citeauthoryear{{Coutu}, {Dufour}, {Bergeron}, {Blouin},
  {Loranger}, {Allard}  \& {Dunlap}}{{Coutu} et~al.}{2019}]{coutuetal19-1}
{Coutu} S.,  {Dufour} P.,  {Bergeron} P.,  {Blouin} S.,  {Loranger} E.,
  {Allard} N.~F.,   {Dunlap} B.~H.,  2019, \mn@doi [ApJ]
  {10.3847/1538-4357/ab46b9}, \href
  {https://ui.adsabs.harvard.edu/abs/2019ApJ...885...74C} {885, 74}

\bibitem[\protect\citeauthoryear{{Cunningham} et~al.,}{{Cunningham}
  et~al.}{2021}]{cunninghametal21-1}
{Cunningham} T.,  et~al., 2021, \mn@doi [MNRAS] {10.1093/mnras/stab553}, \href
  {https://ui.adsabs.harvard.edu/abs/2021MNRAS.503.1646C} {503, 1646}

\bibitem[\protect\citeauthoryear{{Cunto}, {Mendoza}, {Ochsenbein}  \&
  {Zeippen}}{{Cunto} et~al.}{1993}]{cuntoetal93-1}
{Cunto} W.,  {Mendoza} C.,  {Ochsenbein} F.,   {Zeippen} C.~J.,  1993, A\&A,
  \href {1993A&A...275L...5C} {275, L5}

\bibitem[\protect\citeauthoryear{{Debes} \& {Sigurdsson}}{{Debes} \&
  {Sigurdsson}}{2002}]{debes+sigurdsson02-1}
{Debes} J.~H.,  {Sigurdsson} S.,  2002, \mn@doi [ApJ] {10.1086/340291}, \href
  {2002ApJ...572..556D} {572, 556}

\bibitem[\protect\citeauthoryear{{Duncan} \& {Lissauer}}{{Duncan} \&
  {Lissauer}}{1998}]{duncan+lissauer98-1}
{Duncan} M.~J.,  {Lissauer} J.~J.,  1998, \mn@doi [Icarus]
  {10.1006/icar.1998.5962}, \href {1998Icar..134..303D} {134, 303}

\bibitem[\protect\citeauthoryear{{Farihi}, {G{\"a}nsicke}, {Steele}, {Girven},
  {Burleigh}, {Breedt}  \& {Koester}}{{Farihi} et~al.}{2012}]{farihietal12-1}
{Farihi} J.,  {G{\"a}nsicke} B.~T.,  {Steele} P.~R.,  {Girven} J.,  {Burleigh}
  M.~R.,  {Breedt} E.,   {Koester} D.,  2012, \mn@doi [MNRAS]
  {10.1111/j.1365-2966.2012.20421.x}, \href {2012MNRAS.421.1635F} {421, 1635}

\bibitem[\protect\citeauthoryear{{Farihi}, {G{\"a}nsicke}  \&
  {Koester}}{{Farihi} et~al.}{2013}]{farihietal13-1}
{Farihi} J.,  {G{\"a}nsicke} B.~T.,   {Koester} D.,  2013, \mn@doi [Science]
  {10.1126/science.1239447}, \href
  {http://adsabs.harvard.edu/abs/2013Sci...342..218F} {342, 218}

\bibitem[\protect\citeauthoryear{{G{\"a}nsicke}, {Marsh}, {Southworth}  \&
  {Rebassa-Mansergas}}{{G{\"a}nsicke} et~al.}{2006}]{gaensickeetal06-3}
{G{\"a}nsicke} B.~T.,  {Marsh} T.~R.,  {Southworth} J.,   {Rebassa-Mansergas}
  A.,  2006, \mn@doi [Science] {10.1126/science.1135033}, \href
  {2006Sci...314.1908G} {314, 1908}

\bibitem[\protect\citeauthoryear{{G{\"a}nsicke}, {Koester}, {Farihi}, {Girven},
  {Parsons}  \& {Breedt}}{{G{\"a}nsicke} et~al.}{2012}]{gaensickeetal12-1}
{G{\"a}nsicke} B.~T.,  {Koester} D.,  {Farihi} J.,  {Girven} J.,  {Parsons}
  S.~G.,   {Breedt} E.,  2012, \mn@doi [MNRAS]
  {10.1111/j.1365-2966.2012.21201.x}, \href {2012MNRAS.424..333G} {424, 333}

\bibitem[\protect\citeauthoryear{{G{\"a}nsicke}, {Schreiber}, {Toloza},
  {Fusillo}, {Koester}  \& {Manser}}{{G{\"a}nsicke}
  et~al.}{2019}]{gaensickeetal19-1}
{G{\"a}nsicke} B.~T.,  {Schreiber} M.~R.,  {Toloza} O.,  {Fusillo} N. P.~G.,
  {Koester} D.,   {Manser} C.~J.,  2019, \mn@doi [Nat]
  {10.1038/s41586-019-1789-8}, \href
  {https://ui.adsabs.harvard.edu/abs/2019Natur.576...61G} {576, 61}

\bibitem[\protect\citeauthoryear{{Gentile Fusillo}, {G{\"a}nsicke}, {Farihi},
  {Koester}, {Schreiber}  \& {Pala}}{{Gentile Fusillo}
  et~al.}{2017}]{gentileetal17-1}
{Gentile Fusillo} N.~P.,  {G{\"a}nsicke} B.~T.,  {Farihi} J.,  {Koester} D.,
  {Schreiber} M.~R.,   {Pala} A.~F.,  2017, \mn@doi [MNRAS]
  {10.1093/mnras/stx468}, \href
  {http://adsabs.harvard.edu/abs/2017MNRAS.468..971G} {468, 971}

\bibitem[\protect\citeauthoryear{{Gentile Fusillo} et~al.,}{{Gentile Fusillo}
  et~al.}{2021}]{gentilefusilloetal21-1}
{Gentile Fusillo} N.~P.,  et~al., 2021, \mn@doi [MNRAS]
  {10.1093/mnras/stab992}, \href
  {https://ui.adsabs.harvard.edu/abs/2021MNRAS.504.2707G} {504, 2707}

\bibitem[\protect\citeauthoryear{{Girven}, {Brinkworth}, {Farihi},
  {G{\"a}nsicke}, {Hoard}, {Marsh}  \& {Koester}}{{Girven}
  et~al.}{2012}]{girvenetal12-1}
{Girven} J.,  {Brinkworth} C.~S.,  {Farihi} J.,  {G{\"a}nsicke} B.~T.,  {Hoard}
  D.~W.,  {Marsh} T.~R.,   {Koester} D.,  2012, \mn@doi [ApJ]
  {10.1088/0004-637X/749/2/154}, \href {2012ApJ...749..154G} {749, 154}

\bibitem[\protect\citeauthoryear{{Graham}, {Matthews}, {Neugebauer}  \&
  {Soifer}}{{Graham} et~al.}{1990}]{grahametal90-1}
{Graham} J.~R.,  {Matthews} K.,  {Neugebauer} G.,   {Soifer} B.~T.,  1990,
  \mn@doi [ApJ] {10.1086/168907}, \href {1990ApJ...357..216G} {357, 216}

\bibitem[\protect\citeauthoryear{{Green} et~al.,}{{Green}
  et~al.}{2018}]{greenetal18-1}
{Green} G.~M.,  et~al., 2018, \mn@doi [\mnras] {10.1093/mnras/sty1008}, \href
  {https://ui.adsabs.harvard.edu/abs/2018MNRAS.478..651G} {478, 651}

\bibitem[\protect\citeauthoryear{{Green}, {Schlafly}, {Zucker}, {Speagle}  \&
  {Finkbeiner}}{{Green} et~al.}{2019}]{greenetal19-1}
{Green} G.~M.,  {Schlafly} E.,  {Zucker} C.,  {Speagle} J.~S.,   {Finkbeiner}
  D.,  2019, \mn@doi [\apj] {10.3847/1538-4357/ab5362}, \href
  {https://ui.adsabs.harvard.edu/abs/2019ApJ...887...93G} {887, 93}

\bibitem[\protect\citeauthoryear{{Guidry} et~al.,}{{Guidry}
  et~al.}{2021}]{guidryetal21-1}
{Guidry} J.~A.,  et~al., 2021, \mn@doi [ApJ] {10.3847/1538-4357/abee68}, \href
  {https://ui.adsabs.harvard.edu/abs/2021ApJ...912..125G} {912, 125}

\bibitem[\protect\citeauthoryear{{Harrison}, {Bonsor}, {Kama}, {Buchan},
  {Blouin}  \& {Koester}}{{Harrison} et~al.}{2021}]{harrisonetal21-1}
{Harrison} J. H.~D.,  {Bonsor} A.,  {Kama} M.,  {Buchan} A.~M.,  {Blouin} S.,
  {Koester} D.,  2021, \mn@doi [MNRAS] {10.1093/mnras/stab736}, \href
  {https://ui.adsabs.harvard.edu/abs/2021MNRAS.504.2853H} {504, 2853}

\bibitem[\protect\citeauthoryear{{Hollands}, {Koester}, {Alekseev}, {Herbert}
  \& {G{\"a}nsicke}}{{Hollands} et~al.}{2017}]{hollandsetal17-1}
{Hollands} M.~A.,  {Koester} D.,  {Alekseev} V.,  {Herbert} E.~L.,
  {G{\"a}nsicke} B.~T.,  2017, \mn@doi [MNRAS] {10.1093/mnras/stx250}, \href
  {http://adsabs.harvard.edu/abs/2017MNRAS.467.4970H} {467, 4970}

\bibitem[\protect\citeauthoryear{{Hollands}, {G{\"a}nsicke}  \&
  {Koester}}{{Hollands} et~al.}{2018}]{hollandsetal18-1}
{Hollands} M.~A.,  {G{\"a}nsicke} B.~T.,   {Koester} D.,  2018, \mn@doi [MNRAS]
  {10.1093/mnras/sty592}, \href
  {http://adsabs.harvard.edu/abs/2018MNRAS.477...93H} {477, 93}

\bibitem[\protect\citeauthoryear{{Hollands}, {Tremblay}, {G{\"a}nsicke},
  {Koester}  \& {Gentile-Fusillo}}{{Hollands} et~al.}{2021}]{hollandsetal21-1}
{Hollands} M.~A.,  {Tremblay} P.-E.,  {G{\"a}nsicke} B.~T.,  {Koester} D.,
  {Gentile-Fusillo} N.~P.,  2021, \mn@doi [Nat Astron.]
  {10.1038/s41550-020-01296-7}, \href
  {https://ui.adsabs.harvard.edu/abs/2021NatAs...5..451H} {5, 451}

\bibitem[\protect\citeauthoryear{{Horne}}{{Horne}}{1986}]{horne86-1}
{Horne} K.,  1986, PASP, \href {1986PASP...98..609H} {98, 609}

\bibitem[\protect\citeauthoryear{{Hoskin} et~al.,}{{Hoskin}
  et~al.}{2020}]{hoskinetal20-1}
{Hoskin} M.~J.,  et~al., 2020, \mn@doi [MNRAS] {10.1093/mnras/staa2717}, \href
  {https://ui.adsabs.harvard.edu/abs/2020MNRAS.499..171H} {499, 171}

\bibitem[\protect\citeauthoryear{{Izquierdo}, {Toloza}, {G{\"a}nsicke},
  {Rodr{\'\i}guez-Gil}, {Farihi}, {Koester}, {Guo}  \& {Redfield}}{{Izquierdo}
  et~al.}{2021}]{izquierdoetal21-1}
{Izquierdo} P.,  {Toloza} O.,  {G{\"a}nsicke} B.~T.,  {Rodr{\'\i}guez-Gil} P.,
  {Farihi} J.,  {Koester} D.,  {Guo} J.,   {Redfield} S.,  2021, \mn@doi
  [MNRAS] {10.1093/mnras/staa3987}, \href
  {https://ui.adsabs.harvard.edu/abs/2021MNRAS.501.4276I} {501, 4276}

\bibitem[\protect\citeauthoryear{{Jura}}{{Jura}}{2003}]{jura03-1}
{Jura} M.,  2003, \mn@doi [ApJ Lett.] {10.1086/374036}, \href
  {2003ApJ...584L..91J} {584, L91}

\bibitem[\protect\citeauthoryear{{Kaiser}, {Clemens}, {Blouin}, {Dufour},
  {Hegedus}, {Reding}  \& {B{\'e}dard}}{{Kaiser} et~al.}{2020}]{kaiseretal20-1}
{Kaiser} B.~C.,  {Clemens} J.~C.,  {Blouin} S.,  {Dufour} P.,  {Hegedus} R.~J.,
   {Reding} J.~S.,   {B{\'e}dard} A.,  2020, \mn@doi [Sci]
  {10.1126/science.abd1714}, \href
  {https://ui.adsabs.harvard.edu/abs/2020Sci...370.1714K} {370, abd1714}

\bibitem[\protect\citeauthoryear{{Klein}, {Doyle}, {Zuckerman}, {Dufour},
  {Blouin}, {Melis}, {Weinberger}  \& {Young}}{{Klein}
  et~al.}{2021}]{kleinetal21-1}
{Klein} B.~L.,  {Doyle} A.~E.,  {Zuckerman} B.,  {Dufour} P.,  {Blouin} S.,
  {Melis} C.,  {Weinberger} A.~J.,   {Young} E.~D.,  2021, \mn@doi [ApJ]
  {10.3847/1538-4357/abe40b}, \href
  {https://ui.adsabs.harvard.edu/abs/2021ApJ...914...61K} {914, 61}

\bibitem[\protect\citeauthoryear{{Koester}}{{Koester}}{2010}]{koester10-1}
{Koester} D.,  2010, Memorie della Societa Astronomica Italiana,, \href
  {2010MmSAI..81..921K} {81, 921}

\bibitem[\protect\citeauthoryear{{Koester} \& {Kepler}}{{Koester} \&
  {Kepler}}{2019}]{koester+kepler19-1}
{Koester} D.,  {Kepler} S.~O.,  2019, \mn@doi [A\&A]
  {10.1051/0004-6361/201935946}, \href
  {https://ui.adsabs.harvard.edu/abs/2019A&A...628A.102K} {628, A102}

\bibitem[\protect\citeauthoryear{{Koester}, {G{\"a}nsicke}  \&
  {Farihi}}{{Koester} et~al.}{2014}]{koesteretal14-1}
{Koester} D.,  {G{\"a}nsicke} B.~T.,   {Farihi} J.,  2014, \mn@doi [A\&A]
  {10.1051/0004-6361/201423691}, \href
  {http://adsabs.harvard.edu/abs/2014A%26A...566A..34K} {566, A34}

\bibitem[\protect\citeauthoryear{{Koester}, {Kepler}  \& {Irwin}}{{Koester}
  et~al.}{2020}]{koesteretal20-1}
{Koester} D.,  {Kepler} S.~O.,   {Irwin} A.~W.,  2020, \mn@doi [A\&A]
  {10.1051/0004-6361/202037530}, \href
  {https://ui.adsabs.harvard.edu/abs/2020A&A...635A.103K} {635, A103}

\bibitem[\protect\citeauthoryear{{Kramida}, {Ralchenko}, {Reader}  \& {NIST ASD
  Team}}{{Kramida} et~al.}{2016}]{kramidaetal16-1}
{Kramida} A.,  {Ralchenko} Y.,  {Reader} J.,   {NIST ASD Team} 2016, NIST
  Atomic Spectra Database (version 5.4), [Online], \url
  {http://physics.nist.gov/asd}

\bibitem[\protect\citeauthoryear{{Manser} et~al.,}{{Manser}
  et~al.}{2019}]{manseretal19-1}
{Manser} C.~J.,  et~al., 2019, \mn@doi [Sci] {10.1126/science.aat5330}, \href
  {https://ui.adsabs.harvard.edu/abs/2019Sci...364...66M} {364, 66}

\bibitem[\protect\citeauthoryear{{McCleery} et~al.,}{{McCleery}
  et~al.}{2020}]{mccleeryetal20-1}
{McCleery} J.,  et~al., 2020, \mn@doi [MNRAS] {10.1093/mnras/staa2030}, \href
  {https://ui.adsabs.harvard.edu/abs/2020MNRAS.499.1890M} {499, 1890}

\bibitem[\protect\citeauthoryear{{McCord} \& {Sotin}}{{McCord} \&
  {Sotin}}{2005}]{mccord+sotin05-1}
{McCord} T.~B.,  {Sotin} C.,  2005, \mn@doi [J. Geophys. Res.]
  {10.1029/2004JE002244}, \href
  {https://ui.adsabs.harvard.edu/abs/2005JGRE..110.5009M} {110, E05009}

\bibitem[\protect\citeauthoryear{{Melis} et~al.,}{{Melis}
  et~al.}{2012}]{melisetal12-1}
{Melis} C.,  et~al., 2012, \mn@doi [ApJ Lett.] {10.1088/2041-8205/751/1/L4},
  \href {http://adsabs.harvard.edu/abs/2012ApJ...751L...4M} {751, L4}

\bibitem[\protect\citeauthoryear{{Mustill}, {Veras}  \& {Villaver}}{{Mustill}
  et~al.}{2014}]{mustilletal14-1}
{Mustill} A.~J.,  {Veras} D.,   {Villaver} E.,  2014, \mn@doi [MNRAS]
  {10.1093/mnras/stt1973}, \href
  {http://adsabs.harvard.edu/abs/2014MNRAS.437.1404M} {437, 1404}

\bibitem[\protect\citeauthoryear{{Paquette}, {Pelletier}, {Fontaine}  \&
  {Michaud}}{{Paquette} et~al.}{1986}]{paquetteetal86-2}
{Paquette} C.,  {Pelletier} C.,  {Fontaine} G.,   {Michaud} G.,  1986, \mn@doi
  [ApJS] {10.1086/191112}, \href
  {http://adsabs.harvard.edu/abs/1986ApJS...61..197P} {61, 197}

\bibitem[\protect\citeauthoryear{{Provencal}, {Shipman}, {Koester}, {Wesemael}
  \& {Bergeron}}{{Provencal} et~al.}{2002}]{provencaletal02-1}
{Provencal} J.~L.,  {Shipman} H.~L.,  {Koester} D.,  {Wesemael} F.,
  {Bergeron} P.,  2002, \mn@doi [ApJ] {10.1086/338769}, \href
  {2002ApJ...568..324P} {568, 324}

\bibitem[\protect\citeauthoryear{{Raddi}, {G{\"a}nsicke}, {Koester}, {Farihi},
  {Hermes}, {Scaringi}, {Breedt}  \& {Girven}}{{Raddi}
  et~al.}{2015}]{raddietal15-1}
{Raddi} R.,  {G{\"a}nsicke} B.~T.,  {Koester} D.,  {Farihi} J.,  {Hermes}
  J.~J.,  {Scaringi} S.,  {Breedt} E.,   {Girven} J.,  2015, \mn@doi [MNRAS]
  {10.1093/mnras/stv701}, \href
  {http://adsabs.harvard.edu/abs/2015MNRAS.450.2083R} {450, 2083}

\bibitem[\protect\citeauthoryear{{Sackmann}, {Boothroyd}  \&
  {Kraemer}}{{Sackmann} et~al.}{1993}]{sackmannetal93-1}
{Sackmann} I.-J.,  {Boothroyd} A.~I.,   {Kraemer} K.~E.,  1993, \mn@doi [ApJ]
  {10.1086/173407}, \href {1993ApJ...418..457S} {418, 457}

\bibitem[\protect\citeauthoryear{{Schatzman}}{{Schatzman}}{1949}]{schatzman49-1}
{Schatzman} E.,  1949, \mn@doi [ApJ] {10.1086/145202}, \href
  {http://adsabs.harvard.edu/abs/1949ApJ...110..261S} {110, 261}

\bibitem[\protect\citeauthoryear{{Schr{\"o}der} \& {Connon
  Smith}}{{Schr{\"o}der} \& {Connon Smith}}{2008}]{schroeder+connon08-1}
{Schr{\"o}der} K.-P.,  {Connon Smith} R.,  2008, \mn@doi [MNRAS]
  {10.1111/j.1365-2966.2008.13022.x}, \href
  {http://adsabs.harvard.edu/abs/2008MNRAS.386..155S} {386, 155}

\bibitem[\protect\citeauthoryear{{Swan}, {Farihi}, {Koester}, {Holland s},
  {Parsons}, {Cauley}, {Redfield}  \& {G{\"a}nsicke}}{{Swan}
  et~al.}{2019}]{swanetal19-1}
{Swan} A.,  {Farihi} J.,  {Koester} D.,  {Holland s} M.,  {Parsons} S.,
  {Cauley} P.~W.,  {Redfield} S.,   {G{\"a}nsicke} B.~T.,  2019, \mn@doi
  [MNRAS] {10.1093/mnras/stz2337}, \href
  {https://ui.adsabs.harvard.edu/abs/2019MNRAS.490..202S} {490, 202}

\bibitem[\protect\citeauthoryear{{Turner} \& {Wyatt}}{{Turner} \&
  {Wyatt}}{2020}]{turner+wyatt20-1}
{Turner} S. G.~D.,  {Wyatt} M.~C.,  2020, \mn@doi [MNRAS]
  {10.1093/mnras/stz3191}, \href
  {https://ui.adsabs.harvard.edu/abs/2020MNRAS.491.4672T} {491, 4672}

\bibitem[\protect\citeauthoryear{{Vanderbosch} et~al.,}{{Vanderbosch}
  et~al.}{2019}]{vanderboschetal19-1}
{Vanderbosch} Z.,  et~al., 2019, arXiv e-prints, \href
  {https://ui.adsabs.harvard.edu/abs/2019arXiv190809839V} {p. arXiv:1908.09839}

\bibitem[\protect\citeauthoryear{{Vanderbosch} et~al.,}{{Vanderbosch}
  et~al.}{2021}]{vanderboschetal21-1}
{Vanderbosch} Z.~P.,  et~al., 2021, arXiv e-prints, \href
  {https://ui.adsabs.harvard.edu/abs/2021arXiv210602659V} {p. arXiv:2106.02659}

\bibitem[\protect\citeauthoryear{{Vanderburg} et~al.,}{{Vanderburg}
  et~al.}{2015}]{vanderburgetal15-1}
{Vanderburg} A.,  et~al., 2015, \mn@doi [Nat] {10.1038/nature15527}, \href
  {http://adsabs.harvard.edu/abs/2015Natur.526..546V} {526, 546}

\bibitem[\protect\citeauthoryear{{Vanderburg} et~al.,}{{Vanderburg}
  et~al.}{2020}]{vanderburgetal20-1}
{Vanderburg} A.,  et~al., 2020, \mn@doi [Nat] {10.1038/s41586-020-2713-y},
  \href {https://ui.adsabs.harvard.edu/abs/2020Natur.585..363V} {585, 363}

\bibitem[\protect\citeauthoryear{{Veras} \& {G{\"a}nsicke}}{{Veras} \&
  {G{\"a}nsicke}}{2015}]{veras+gaensicke15-1}
{Veras} D.,  {G{\"a}nsicke} B.~T.,  2015, \mn@doi [MNRAS]
  {10.1093/mnras/stu2475}, \href
  {http://adsabs.harvard.edu/abs/2015MNRAS.447.1049V} {447, 1049}

\bibitem[\protect\citeauthoryear{{Veras}, {Mustill}, {G{\"a}nsicke},
  {Redfield}, {Georgakarakos}, {Bowler}  \& {Lloyd}}{{Veras}
  et~al.}{2016}]{verasetal16-1}
{Veras} D.,  {Mustill} A.~J.,  {G{\"a}nsicke} B.~T.,  {Redfield} S.,
  {Georgakarakos} N.,  {Bowler} A.~B.,   {Lloyd} M.~J.~S.,  2016, \mn@doi
  [MNRAS] {10.1093/mnras/stw476}, \href
  {http://adsabs.harvard.edu/abs/2016MNRAS.458.3942V} {458, 3942}

\bibitem[\protect\citeauthoryear{{Villaver} \& {Livio}}{{Villaver} \&
  {Livio}}{2007}]{villaver+livio07-1}
{Villaver} E.,  {Livio} M.,  2007, \mn@doi [ApJ] {10.1086/516746}, \href
  {2007ApJ...661.1192V} {661, 1192}

\bibitem[\protect\citeauthoryear{{Walkup}, {Stewart}  \& {Pritchard}}{{Walkup}
  et~al.}{1984}]{walkupetal84-1}
{Walkup} R.,  {Stewart} B.,   {Pritchard} D.~E.,  1984, Phys. Rev. A, 29, 169

\bibitem[\protect\citeauthoryear{{Wilson}, {G{\"a}nsicke}, {Koester}, {Toloza},
  {Pala}, {Breedt}  \& {Parsons}}{{Wilson} et~al.}{2015}]{wilsonetal15-1}
{Wilson} D.~J.,  {G{\"a}nsicke} B.~T.,  {Koester} D.,  {Toloza} O.,  {Pala}
  A.~F.,  {Breedt} E.,   {Parsons} S.~G.,  2015, \mn@doi [MNRAS]
  {10.1093/mnras/stv1201}, \href
  {http://adsabs.harvard.edu/abs/2015MNRAS.451.3237W} {451, 3237}

\bibitem[\protect\citeauthoryear{{Wilson}, {G{\"a}nsicke}, {Farihi}  \&
  {Koester}}{{Wilson} et~al.}{2016}]{wilsonetal16-1}
{Wilson} D.~J.,  {G{\"a}nsicke} B.~T.,  {Farihi} J.,   {Koester} D.,  2016,
  \mn@doi [MNRAS] {10.1093/mnras/stw844}, \href
  {https://ui.adsabs.harvard.edu/abs/2016MNRAS.459.3282W} {459, 3282}

\bibitem[\protect\citeauthoryear{{Wolff}, {Koester}  \& {Liebert}}{{Wolff}
  et~al.}{2002}]{wolffetal02-1}
{Wolff} B.,  {Koester} D.,   {Liebert} J.,  2002, \mn@doi [A\&A]
  {10.1051/0004-6361:20020194}, \href
  {https://ui.adsabs.harvard.edu/abs/2002A&A...385..995W} {385, 995}

\bibitem[\protect\citeauthoryear{{Xu}, {Zuckerman}, {Dufour}, {Young}, {Klein}
  \& {Jura}}{{Xu} et~al.}{2017}]{xuetal17-1}
{Xu} S.,  {Zuckerman} B.,  {Dufour} P.,  {Young} E.~D.,  {Klein} B.,   {Jura}
  M.,  2017, \mn@doi [ApJ Lett.] {10.3847/2041-8213/836/1/L7}, \href
  {http://adsabs.harvard.edu/abs/2017ApJ...836L...7X} {836, L7}

\bibitem[\protect\citeauthoryear{{Zuckerman} \& {Becklin}}{{Zuckerman} \&
  {Becklin}}{1987}]{zuckerman+becklin87-1}
{Zuckerman} B.,  {Becklin} E.~E.,  1987, \mn@doi [Nat] {10.1038/330138a0},
  \href {1987Natur.330..138Z} {330, 138}

\bibitem[\protect\citeauthoryear{{Zuckerman}, {Koester}, {Reid}  \&
  {H{\"u}nsch}}{{Zuckerman} et~al.}{2003}]{zuckermanetal03-1}
{Zuckerman} B.,  {Koester} D.,  {Reid} I.~N.,   {H{\"u}nsch} M.,  2003, \mn@doi
  [ApJ] {10.1086/377492}, \href {2003ApJ...596..477Z} {596, 477}

\bibitem[\protect\citeauthoryear{{Zuckerman}, {Koester}, {Melis}, {Hansen}  \&
  {Jura}}{{Zuckerman} et~al.}{2007}]{zuckermanetal07-1}
{Zuckerman} B.,  {Koester} D.,  {Melis} C.,  {Hansen} B.~M.,   {Jura} M.,
  2007, \mn@doi [ApJ] {10.1086/522223}, \href {2007ApJ...671..872Z} {671, 872}

\bibitem[\protect\citeauthoryear{{Zuckerman}, {Melis}, {Klein}, {Koester}  \&
  {Jura}}{{Zuckerman} et~al.}{2010}]{zuckermanetal10-1}
{Zuckerman} B.,  {Melis} C.,  {Klein} B.,  {Koester} D.,   {Jura} M.,  2010,
  \mn@doi [ApJ] {10.1088/0004-637X/722/1/725}, \href {2010ApJ...722..725Z}
  {722, 725}

\bibitem[\protect\citeauthoryear{{Zuckerman}, {Koester}, {Dufour}, {Melis},
  {Klein}  \& {Jura}}{{Zuckerman} et~al.}{2011}]{zuckermanetal11-1}
{Zuckerman} B.,  {Koester} D.,  {Dufour} P.,  {Melis} C.,  {Klein} B.,   {Jura}
  M.,  2011, \mn@doi [ApJ] {10.1088/0004-637X/739/2/101}, \href
  {2011ApJ...739..101Z} {739, 101}

\makeatother
\end{thebibliography}




%


\bsp	
\label{lastpage}
\end{document}